\documentclass{article}

\usepackage{parskip}
\usepackage[margin = 1.5in]{geometry}
\usepackage{times}
\usepackage{graphicx} 
\usepackage{subfigure} 

\usepackage{amsmath}
\usepackage{amssymb}
\usepackage{amsfonts}
\usepackage{mathtools}
\usepackage{graphicx}
\usepackage{color}
\usepackage{epstopdf}
\usepackage{amsthm}

\newtheorem{theorem}{Theorem}
\newtheorem{lemma}[theorem]{Lemma}

\newcommand{\argmax}{\operatornamewithlimits{argmax}}
\newcommand\abs[1]{\left|#1\right|}
\newcommand{\norm}[1]{\left\lVert#1\right\rVert}
\newcommand{\pn}[1]{p_n\left({#1}\right)}

\usepackage[numbers]{natbib}

\usepackage{algorithm,caption}
\usepackage{algorithmic}

\usepackage{booktabs}       
\usepackage{multirow}
\usepackage{amssymb}
\usepackage{amsmath}
\usepackage{mathtools}
\usepackage{hyperref}

\begin{document}
\title{Faster Greedy MAP Inference for Determinantal Point Processes}
\author{
Insu Han\thanks{School of Electrical Engineering, Korea Advanced Institute of Science and Technology, Korea. 
Emails: \{hawki17, kyoungsoo, jinwoos\}@kaist.ac.kr
} \and 
Prabhanjan Kambadur\thanks{
Bloomberg LP, 731 Lexington Avenue, New York, NY, 10069.
Email: prabhanjankambadur@gmail.com
} \and 
Kyoungsoo Park$^*$ 
\and
Jinwoo Shin$^*$ 
}

\maketitle

\begin{abstract} 
Determinantal point processes (DPPs) are popular probabilistic models that arise in many
machine learning tasks, where distributions of diverse sets are characterized by matrix determinants.
In this paper, we develop fast algorithms to find the most likely configuration (MAP)
of large-scale DPPs, which is NP-hard in general.
Due to the submodular nature of the MAP objective, greedy algorithms have been used
with empirical success.
Greedy implementations require computation of log-determinants, matrix inverses
or solving linear systems at each iteration.
We present faster implementations of the greedy algorithms by utilizing the
complementary benefits of two log-determinant approximation schemes:
(a) first-order expansions to the matrix log-determinant function and (b) 
high-order expansions to the scalar log function with stochastic trace estimators.
In our experiments, our algorithms are significantly faster than their
competitors for large-scale instances, while sacrificing marginal accuracy.
\end{abstract}

\section{Introduction} \label{sec:intro}

Determinantal point processes (DPPs) are elegant probabilistic models, first introduced by \cite{macchi1975coincidence}, who called them `fermion processes'.
Since then, DPPs have been extensively studied in the fields of quantum physics
and random matrices \cite{johansson2006course}, giving rise to a beautiful theory \cite{daley2007introduction}.
The characteristic of DPPs is repulsive behavior, which makes them useful for modeling diversity.
%

Recently, they have been applied in many machine learning tasks such as
summarization \cite{gong2014diverse}, human pose detection \cite{kulesza2012determinantal},
clustering \cite{kang2013fast} and tweet time-line generation \cite{yao2016tweet}.
In particular, their computational advantage compared to other probabilistic models
is that many important inference tasks are computationally tractable.
For example, conditioning, sampling \cite{kang2013fast} and marginalization of DPPs
admit polynomial-time/efficient algorithms, while those on popular graphical
models \cite{jordan1998learning} do not, i.e., they are NP-hard. 
One exception is the MAP inference (finding the most likely configuration), which is our
main interest; the MAP computation is known to be NP-hard even for DPPs \cite{kulesza2012determinantal}.

The distribution of diverse sets under DPPs is characterized by 
determinants of submatrices formed by their features, 
and the corresponding MAP inference reduces to finding a submatrix that
maximizes its determinant.
It is well known that the matrix log-determinant is a submodular function; that is,
the MAP inference of DPPs is a special instance of submodular maximization \cite{kulesza2012determinantal}.
Greedy algorithms have been shown to have the best worst-case approximation guarantees
for many instances of submodular maximization; for example, $(1-1/e)$-approximation for
monotone functions. 
Furthermore, it has been often empirically observed that greedy algorithms provide 
near optimal solutions \cite{krause2008near}.
Hence, greedy algorithms have been also applied for the DPP
task \cite{kulesza2012determinantal,yao2016tweet,zhang2016block}.
Known implementations of greedy selection on DPP require computation of log-determinants, 
matrix inversions \cite{kulesza2012determinantal} or solving linear systems \cite{li2016gaussian}.
Consequently, they run in {$O(d^4)$} time where $d$ is the total number of items
(see Section \ref{sec:naiveimp}). In this paper, we propose faster greedy implementations
that run in {$O(d^3)$} time.

{\bf Contribution.}
%
Our high-level idea is to amortize greedy operations by utilizing log-determinant
approximation schemes.
A greedy selection requires computation of marginal gains of log-determinants; we
consider their first-order (linear) approximations.
We observe that the computation of multiple marginal gains can be amortized into a
single run of a linear solver, 
in addition to multiple vector inner products.
We choose the popular conjugate gradient descent ($\mathtt{CG}$) \cite{saad2003iterative}
as a linear solver. 
In addition, for improving the quality of first-order approximations, we partition remaining
items into $p\geq 1$ sets (via some clustering algorithm), and apply the first-order approximations
in each partition.
The resulting approximate computation of multiple marginal gains at each greedy selection
requires $2p$ runs of $\mathtt{CG}$ under the Schur complement, and the overall running time
of the proposed greedy algorithm becomes {$O(d^3)$} under the choice of
$p=O(1)$ (see Section \ref{sec:greedy}). 

Next, for larger-scale DPPs, we develop an even faster greedy algorithm using a batch strategy.
%
%
In addition to using the first-order approximations of log-determinants under a
partitioning scheme, we add $k>1$ elements instead of a single element to the
current set, where we sample some candidates among all possible $k$ elements
to relax the expensive cost of computing all marginal gains.
Intuitively, the random batch selection makes the algorithm $k$ times faster, while potentially
hurting the approximation quality.
Now, we suggest running the recent fast log-determinant approximation scheme ($\mathtt{LDAS}$)
\cite{han2015large} $p$ times, instead of running $\mathtt{CG}$ $pk$ times under the Schur
complement, where $\mathtt{LDAS}$ utilizes high-order, i.e., polynomial, approximations 
to the scalar log function with stochastic trace estimators. 
Since the complexities of running $\mathtt{LDAS}$ and $\mathtt{CG}$ are comparable, running
the former $p$ times is faster than running the latter $pk$ times if $k>1$.

Finally, we discovered a novel scheme for boosting the approximation quality by 
sharing random vectors among many runs of $\mathtt{LDAS}$, and also establish
theoretical justification why this helps.
Our experiments
on 
both synthetic and real-world dataset show that
the proposed algorithms are significantly faster than competitors for large-scale instances, 
while losing marginal approximation ratio. 

{\bf Related work.} 
To the best of our knowledge,
this is the first work that aims for developing faster greedy algorithms
specialized for the MAP inference of DPP, 
while
there has been several efforts on those for
general submodular maximization.
An accelerated greedy algorithm, called lazy evaluation, was first proposed by \cite{minoux1978accelerated} 
which maintains the upper bounds on the marginal gains instead of recomputing exact values.
In each iteration, only elements with the maximal bound compute the exact gain,
which still bounds on the exact value due to submodularity.
For the DPP case, we also observe that the lazy algorithm 
is significantly faster than the standard greedy one, while the outputs of both are 
equal. Hence, we compare our algorithms
with the lazy one (see Section \ref{sec:exp}).



Another natural approach is on stochastic greedy selections 
computing marginal gains of randomly selected elements.
Its worst-case approximation guarantee was also studied \cite{mirzasoleiman2015lazier},
under the standard, non-batch, greedy algorithm.
The idea of stochastic selections 
can be also applied to our algorithms, where
we indeed apply it
for designing our faster batch greedy algorithm as mentioned earlier.
Recently, \cite{buchbinder2015tight} proposed a `one-pass' greedy algorithm 
where each greedy selection requires computing only a single marginal gain,
i.e., the number of marginal gains necessary to compute can be significantly reduced.
However, this algorithm is attractive only for the case when evaluating a marginal gain
does not increase with respect to the size of the current set, which does not hold
for the DPP case. As reported in Section \ref{sec:exp},
it performs significantly worse than ours in both their approximation qualities 
and running times.

There have been also several efforts to design parallel/distributed implementations of
greedy algorithms: \citep{pan2014parallel} use parallel strategies for 
the above one-pass greedy algorithm and
\citep{kumar2015fast} adapt a MapReduce paradigm 
for implementing greedy algorithms in distributed settings. 
One can also parallelize our algorithms easily since they
require independent runs of matrix-vector (or vector inner) products,
but we do not explore this aspect in this paper.
Finally, we remark that a non-greedy algorithm was studied in 
\cite{gillenwater2012near} for better MAP qualities of DPP,
but it is much slower than ours 
as reported in Section \ref{sec:exp}.

{\bf Organization.}
We introduce the necessary background in Section \ref{sec:prelim}, 
and present the proposed algorithms 
in Section \ref{sec:greedy} and 
Section \ref{sec:batch}.
Proofs and Experimental results are presented in Section \ref{sec:proof} and Section \ref{sec:exp}, respectively.
\section{Preliminaries} \label{sec:prelim}

We start by defining a necessary notation.
Our algorithms for determinantal point processes (DPPs)
select elements from the ground set of $d$ items $\mathcal{Y} = [d] := \{1,2,\dots,d\}$ 
and denote the set of all subsets of $\mathcal{Y}$ by $2^\mathcal{Y}$.
For any positive semidefinite matrix $L \in \mathbb{R}^{d \times d}$,
we denote $\lambda_{\min}$ and $\lambda_{\max}$ to be the smallest and the largest eigenvalues 
of $L$.
Given subset $X,Y \subseteq \mathcal{Y}$, we use $L_{X,Y}$ to denote 
the submatrix of $L$ obtained by entries in rows and
columns indexed by $X$ and $Y$, respectively.
For notational simplicity, we let $L_{X,X} = L_X$ and $L_{X,\{i\}} = L_{X,i}$ for $i \in \mathcal{Y}$.
In addition, $\overline{L}_{X}$ is defined as the average of $L_{X \cup \{ i \}}$ for $ i \in \mathcal{Y} \setminus X$.
Finally, $\left\langle \cdot, \cdot  \right\rangle$ means the matrix/vector inner product or element-wise product sum.

In Section \ref{sec:dpp}, we introduce the {\it maximum a posteriori} (MAP) inference
of DPP, then
the standard greedy optimization scheme and its na\"ive implementations
are described in Section \ref{sec:greedypre}
and Section \ref{sec:naiveimp}, respectively.

\subsection{Determinantal Point Processes}\label{sec:dpp}

DPPs are probabilistic models for subset selection 
of a finite ground set $\mathcal{Y} = [d]$ that captures both quality and diversity.
Formally, it defines the following distribution on $2^{\mathcal Y}$:
for random variable $\mathbf{X} \subseteq \mathcal{Y}$ drawn from given DPP, 
we have
$$
\Pr \left[ \mathbf{X} = X \right] \propto \det \left( L_X \right),
$$
where $L \in \mathbb{R}^{d \times d}$ is a positive definite matrix called an {\it $L$-ensemble} kernel.
Under the distribution,
several probabilistic inference tasks are required for real-world applications,
including MAP \cite{gong2014diverse,gillenwater2012near,yao2016tweet}, sampling \cite{kathuria2016sampling,kang2013fast,li2016efficient}, marginalization and conditioning \cite{gong2014diverse}.
In particular, {we are interested in} the MAP inference, i.e., finding the most diverse subset $Y$ of $\mathcal{Y}$ {that}
achieves the highest probability, i.e., $ \arg\max_{Y\subseteq \mathcal{Y}} \det(L_Y)$,
possibly under some constraints on $Y$.
Unlike other inference tasks on DPP, it is known that MAP is a NP-hard problem \cite{kulesza2012determinantal}.

\subsection{Greedy Submodular Maximization}\label{sec:greedypre}
A set function $f : 2^\mathcal{Y} \rightarrow \mathbb{R}$ is submodular 
if its marginal gains are decreasing, i.e.,
\begin{equation*}
f( X\cup \{ i \}) - f(X) \geq f(Y \cup \{ i \}) - f(Y),
\end{equation*}
for every $X \subseteq Y \subset \mathcal{Y}$ and every $i \in \mathcal{Y} \setminus Y$.
We say $f$ is monotone if $f(X) \leq f(Y)$ for every $X \subseteq Y$.
It is well known that
DPP has the submodular structure, i.e.,
$f = \log \det$ is submodular.

The submodular maximization task
is to find a subset maximizing a submodular function $f$, which
corresponds to the MAP inference task in the DPP case. 
Hence, it is NP-hard and a popular approximate scheme
is the following greedy procedure \cite{nemhauser1978analysis}:
initially, $X \leftarrow \emptyset$ and iteratively update 
$X \leftarrow X \cup \{ i_{\max} \}$ for
\begin{equation}\label{eq:maxmarginalgain}
i_{\max} = \argmax_{i \in \mathcal{Y} \setminus X} f ( {X\cup\{i\}} ) - f(X),
\end{equation}
as long as $f({X\cup\{i_{\max}\}}) > f({X})$.
For the monotone case, 
it guarantees $(1-1/e)$-approximation \cite{nemhauser1978analysis}.
Under
some modifications of the standard greedy procedure,
$2/5$-approximation can be guaranteed even for non-monotone functions \cite{feige2011maximizing}. 
Irrespectively of such theoretical guarantees,
it has been empirically observed that greedy selection \eqref{eq:maxmarginalgain} provides near
optimal solutions in practice \cite{krause2008near, sharma2015greedy, yao2016tweet,zhang2016block}.

\subsection{Na\"ive Implementations of Greedy Algorithm}\label{sec:naiveimp}

%
Log-determinant or related computations, which are at the heart of greedy
algorithms for MAP inference of DPPs, 
are critical to compute the marginal gain
$\log \det L_{X \cup \{ i\}} - \log \det L_X$.
{Since the exact computations of log-determinants} might be slow, i.e., requires $O(d^3)$ time for $d$-dimensional matrices, we introduce recent efficient log-determinant approximation schemes ($\mathtt{LDAS}$).
%
The log-determinant of a symmetric positive definite matrix $A$ 
can be approximated by combining
(a) Chebyshev polynomial expansion of scalar $\log$ function and
(b) matrix trace estimators via Monte Carlo methods:
\begin{align*}
\log \det A = {\tt tr}\left(\log A\right)
\stackrel{(a)}{\approx} {\tt tr}\left( p_n(A)\right)  
\stackrel{(b)}{\approx} \frac1m \sum_{t=1}^m \mathbf{v}^{(t)\top} p_n (A) \mathbf{v}^{(t)}. \end{align*}
Here, $p_n(x)$ is a polynomial expansion of degree $n$ approximating $\log x$
and $\mathbf{v}^{(1)}, \dots, \mathbf{v}^{(m)}$ are random vectors used for estimating the trace of $p_n(A)$.
Several polynomial expansions, including
Taylor~\cite{boutsidis2015randomized}, Chebyshev~\cite{han2015large} and
Legendre~\cite{peng2015large}
have been studied. 
For trace estimation, several random vectors have been also studied 
\cite{avron2011randomized}, e.g.,
the Hutchinson method~\cite{hutchinson1990stochastic} chooses elements of 
$\mathbf{v}$ as i.i.d. random numbers in $\{ -1,+1\}$ so that 
$\mathbf{E}\left[\mathbf{v}^\top A \mathbf{v} \right] = {\tt tr}\left( A \right)$.
In this paper, 
we use 
$\mathtt{LDAS}$ using the Chebyshev polynomial and 
Hutchinson method \cite{han2015large}, but
one can use other alternatives as well. 


\begin{algorithm}[t]
\caption*{Log-determinant Approximation Scheme ($\mathtt{LDAS}$) \cite{han2015large}} \label{alg:logdet}
\begin{algorithmic}
\STATE {\bfseries Input:} symmetric matrix $A \in \mathbb{R}^{d \times d}$ with eigenvalues in $[\delta,1-\delta]$,  sampling number $m$ and polynomial degree $n$
\STATE {\bfseries Initialize:} $\Gamma \leftarrow 0$
     \STATE $c_j  \leftarrow$ $j$-th coefficient of Chebyshev expansion of $\log x$ on $[\delta,1-\delta]$ for $0 \leq j \leq n$.
  \FOR {$i = 1$ { \bfseries to } $m$}
  \STATE Draw a random vector $\mathbf{v}^{(i)} \in \{ -1,+1\}^d$ whose entries are uniformly distributed.
  \STATE $\mathbf{w}_0^{(i)} \leftarrow \mathbf{v}^{(i)}$ and $\mathbf{w}_1^{(i)} \leftarrow \frac{2}{1-2\delta}A\mathbf{v}^{(i)}-\frac{1}{1-2\delta} \mathbf{v}^{(i)}$
  \STATE $\mathbf{u} \leftarrow c_0 \mathbf{w}_0^{(i)} + c_1 \mathbf{w}_1^{(i)}$
  \FOR {$j = 2$ { \bfseries to } $n$}
  \STATE $\mathbf{w}_2^{(i)} \leftarrow \frac{4}{1-2\delta} A \mathbf{w}_1^{(i)} - \frac{2}{1-2\delta} \mathbf{w}_1^{(i)} - \mathbf{w}_0^{(i)}$
  \STATE $\mathbf{u} \leftarrow \mathbf{u} + c_j \ \mathbf{w}_2^{(i)}$
  \STATE $\mathbf{w}_0^{(i)} \leftarrow \mathbf{w}_1^{(i)}$ and $\mathbf{w}_1^{(i)} \leftarrow \mathbf{w}_2^{(i)}$
  \ENDFOR
  \STATE $\Gamma \leftarrow \Gamma + \mathbf{v}^{(i)\top} \mathbf{u}/ m $
  \ENDFOR
\STATE {\bfseries Output:} $\Gamma$
\end{algorithmic}
\end{algorithm}

Observe that $\mathtt{LDAS}$ only requires matrix-vector multiplications 
and its running time is $\Theta\left( d^{2}\right)$ for constants $m,n=O(1)$.
One can directly use $\mathtt{LDAS}$ for computing \eqref{eq:maxmarginalgain}
and the resulting greedy algorithm runs in {$\Theta(d \cdot T_{\mathtt{GR}}^3)$} time
where the number of greedy updates on the current set $X$ is $ T_{\mathtt{GR}}$.
Since $T_{\mathtt{GR}}=O(d)$, the complexity is simply {$O(d^{4})$}.
An alternative way to achieve the same complexity is to use
the Schur complement \cite{ouellette1981schur}: 
\begin{align} \label{eqn:cggain}
\log \det L_{X \cup \{ i\}} - \log \det L_X = \log \left( L_{i,i} - L_{i,X} L_{X}^{-1} L_{X,i}\right).
\end{align}
This requires a linear solver to compute $L_{X}^{-1} L_{X,i}$; 
conjugate gradient descent ($\mathtt{CG}$) \cite{greenbaum1997iterative} is
a popular choice in practice. 
Hence, if one applies $\mathtt{CG}$
to compute the max-marginal gain \eqref{eq:maxmarginalgain},
the resulting greedy algorithm runs in 
{
$\Theta(d \cdot T_{\mathtt{GR}}^{3}\cdot T_{\mathtt{CG}})$ time,
}
where $T_{\mathtt{CG}}$ denotes the number of iterations of each $\mathtt{CG}$ run.
In the worst case, $\mathtt{CG}$ converges to the exact solution 
when $T_{\mathtt{CG}}$ grows with the matrix dimension, but
for practical purposes,
it typically provides a very accurate solution in few iterations, i.e., 
$T_{\mathtt{CG}} = O(1)$.
Recently, Gauss quadrature via Lanczos iteration is used for
efficient computing of $L_{i,X} L_{X}^{-1} L_{X,i}$ \cite{li2016gaussian}.
Although it guarantees rigorous upper/lower bounds, 
$\mathtt{CG}$ is faster and accurate enough for most practical purposes.

In summary, the greedy MAP inference of DPP can be implemented efficiently
via $\mathtt{LDAS}$ or $\mathtt{CG}$.
The faster implementations proposed in this paper smartly {employ both of them} as key components utilizing
their {complementary} benefits. 

\section{Faster Greedy DPP Inference} \label{sec:greedy}

In this section, we provide a faster 
greedy submodular maximization
scheme for the MAP inference of DPP. 
We explain our key ideas in Section
\ref{sec:idea1} and then, provide the formal algorithm description
in Section \ref{sec:algdec1}.

\subsection{Key Ideas}\label{sec:idea1}
\noindent {\bf First-order approximation of log-determinant.}
The main computational bottleneck of a greedy algorithm is to evaluate the marginal gain \eqref{eq:maxmarginalgain}
for every element not in the current set.
To reduce the time complexity, 
we consider the following first-order, i.e., linear, approximation of log-determinant
as:\footnote{
$\nabla_X \log\det X=\left(X^{-1}\right)^\top$
}
\begin{align} 
\argmax_{i\in \mathcal Y\setminus X}
\log \det L_{X \cup \{ i \}} - \log \det {L}_{X}\notag 
& = 
\argmax_{i\in \mathcal Y\setminus X}
\log \det L_{X \cup \{ i \}} - \log \det \overline{L}_{X}  \notag \\
&\approx 
\argmax_{i\in \mathcal Y\setminus X}
\left\langle {\overline{L}_{X}^{-1}}, L_{X \cup \{ i \}} - {\overline{L}_{X}} \right\rangle, 
\label{eqn:linear}
\end{align}
where we recall that $\overline{L}_{X}$ is the average of $L_{X \cup \{ i \}}$.
Observe that
computing \eqref{eqn:linear} requires
the vector inner product of a single column (or row) of $\overline{L}_{X}^{-1}$ and $L_{X \cup \{ i \}} - \overline{L}_{X}$
because $L_{X \cup \{ i \}}$ and $\overline{L}_{X}$ share almost {all} entries except {a} single row and {a} column.

To obtain a single column of $\overline{L}_X^{-1}$, 
one can solve a linear system 
using the $\mathtt{CG}$ algorithm.
More importantly,
it suffices to run $\mathtt{CG}$ once for computing \eqref{eqn:linear}, while
the na\"ive greedy implementation in Section \ref{sec:naiveimp}
has to run $\mathtt{CG}$ $|\mathcal Y\setminus X|$ times.
As we mentioned earlier, after obtaining the single column of  $\overline{L}_{X}^{-1}$ using $\mathtt{CG}$, 
one has to perform $|\mathcal Y\setminus X|$ vector inner products in \eqref{eqn:linear}, 
but it is much cheaper than $|\mathcal Y\setminus X|$ $\mathtt{CG}$ runs requiring matrix-vector multiplications.



\noindent {\bf Partitioning.}
In order to further improve the quality of first-order approximation \eqref{eqn:linear}, 
we partition $ \mathcal{Y} \setminus X$ into $p$ distinct subsets 
so that
\begin{align*}
\| L_{X \cup \{ i \}} -  \overline{L}_{X} \|_F ~~\gg~~ 
\| L_{X \cup \{ i \}} - \overline{L}_{X}^{(j)} \|_F, 
\end{align*}
where an element $i$ is in the partition $j \in \left[p\right]$,
$\overline{L}_{X}^{(j)}$
is the average of $L_{X \cup \{ i \}}$ for $i$ in the partition $j$,
and $\norm{\cdot}_F$ is the Frobenius norm.
Since $L_{X \cup \{ i \}}$ becomes closer to the average $\overline{L}_{X}^{(j)}$,
one can expect that the first-order approximation quality in \eqref{eqn:linear} is improved.
But, we now need a more expensive procedure to approximate the marginal gain:
\begin{align*} 
&\log \det L_{X \cup \{ i \}} - \log \det {L}_{X} \\
&= \left(\log \det L_{X \cup \{ i \}} - \log \det \overline{L}_{X}^{(j)} \right) 
+\left(\log \det \overline{L}_{X}^{(j)} - \log \det {L}_{X}\right)\\
&
\approx 
\underbrace{\left\langle {\left(\overline{L}_{X}^{(j)}\right)^{-1}}, L_{X \cup \{ i \}} - {\overline{L}_{X}^{(j)}} \right\rangle}_{(a)} 
+ \underbrace{\left(\log \det \overline{L}_{X}^{(j)} - \log \det {L}_{X}\right)}_{(b)}.
\end{align*}
The first term (a) can be computed efficiently as we explained earlier,
but we have to run
$\mathtt{CG}$ $p$ times for computing single columns of 
$\overline{L}_{X}^{(1)},\dots,\overline{L}_{X}^{(p)}$.
The second term (b) can be also computed using $\mathtt{CG}$ 
similarly {to} \eqref{eqn:cggain} under the Schur complement.
Hence, one has to run $\mathtt{CG}$ $2p$ times in total.
If $p$ is large, the overall complexity becomes larger,
but the approximation quality {improves as well}.
We also note that one can try various clustering algorithms, e.g., 
$k$-means or Gaussian mixture.
Instead, we use a simple random partitioning scheme
because {it is not only} the fastest {method but it} also works well in our experiments.

\subsection{Algorithm Description and Guarantee}\label{sec:algdec1}

The formal description of the proposed algorithm is described in { Algorithm \ref{alg:glin}}.

\begin{algorithm}[th]
\caption{Faster Greedy DPP Inference} \label{alg:glin}
\begin{algorithmic}[1]
{
\STATE {\bf Input:} kernel matrix $L \in \mathbb{R}^{d \times d}$ and number of partitions $p$
\STATE {\bf Initialize:} $X \leftarrow \emptyset$
\WHILE {$\mathcal{Y} \setminus X\neq \emptyset$}
\STATE Partition $\mathcal{Y} \setminus X$ randomly into $p$ subsets.
\FOR {$j = 1$ { \bfseries to } $p$}
\STATE $\overline{L}_X^{(j)} \leftarrow \text{average of } L_{X \cup \{ i \}}$ for $i$ in the partition $j$
\STATE $\mathbf{z}^{(j)} \leftarrow$
$\left( |X|+1\right)$-th column of $\left(\overline{L}_X^{(j)}\right)^{-1}$ 
\STATE ${\Gamma}_j \leftarrow \log \det \overline{L}_X^{(j)} - \log \det L_X$ 
\ENDFOR
\FOR{$i \in \mathcal{Y} \setminus X$}
\STATE ${\Delta}_i \leftarrow \left\langle L_{X \cup \{ i \}} - {\overline{L}_{X}^{(j)}}, \mathtt{Mat} \left(\mathbf{z}^{(j)}\right) \right\rangle \footnotemark + \Gamma_j$\\
where element $i$ is included in partition $j$.
\ENDFOR
\STATE $i_{\max} \leftarrow \argmax_{i \in \mathcal{Y}\setminus X} {\Delta}_i $
\IF{$\log\det L_{X\cup\{i_{\max}\}} - \log\det L_{X} < 0$}
\STATE {\bf return $X$} 
\ENDIF
\STATE $X \leftarrow X \cup \{ i_{\max} \}$
\ENDWHILE
}
\end{algorithmic}
\end{algorithm}
\footnotetext{For $Z\in \mathbb{R}^{d \times k}$,
$\mathtt{Mat}(Z) \in \mathbb{R}^{d \times d}$ is defined 
whose the last $k$ columns and rows are equal to $Z$ and $Z^\top$, respectively, 
and other entries set to $0$.}

As we explained in Section \ref{sec:idea1},
the lines 7, 8 require to run $\mathtt{CG}$. 
Hence, 
the overall complexity becomes 
$\Theta(T_{\mathtt{GR}}^{3}\cdot T_{\mathtt{CG}} \cdot p + d \cdot T_{\mathtt{GR}}^{2} )
=\Theta(T_{\mathtt{GR}}^{3}+ d \cdot T_{\mathtt{GR}}^{2} )$, 
where we choose
$p,T_{\mathtt{CG}} = O(1)$.
%
Since $T_{\mathtt{GR}}=O(d)$,
it is simply $O(d^{3})$ and
better than the complexity $O(d^4)$ 
of the na\"ive implementations described in Section \ref{sec:naiveimp}.
{In particular, if kernel matrix $L$ is sparse, i.e., number of non-zeros of each column/row is $O(1)$, 
ours has the complexity $\Theta(T_{\mathtt{GR}}^{2}+ d \cdot T_{\mathtt{GR}} )$
while the na\"ive approaches are still worse having the complexity $\Theta(d\cdot T_{\mathtt{GR}}^{2})$.}

We also provide the following approximation guarantee of { Algorithm \ref{alg:glin}}
for the monotone case, where its proof is given in Section \ref{sec:pf:thm:alg1}.

\begin{theorem} \label{thm:alg1}
Suppose the smallest eigenvalue of $L$ is greater than 1.
Then, it holds that
\begin{align*}
&\log \det L_X \geq \left( 1 - 1/e \right) \max_{Z \subseteq \mathcal{Y}, |Z|=|X|} \log \det L_Z - 2|X| \varepsilon. 
\end{align*}
where 
$$\varepsilon = \max_{X \subseteq \mathcal{Y}, i \in \mathcal{Y} \setminus X \atop j \in \left[p\right]}
\abs{
\log \frac{\det L_{X \cup \{ i \}}}{\det \overline{L}_X^{(j)}} - \left\langle \left(\overline{L}_{X}^{(j)}\right)^{-1}, L_{X \cup \{ i \}} - \overline{L}_{X}^{(j)} \right\rangle
}$$
and $X$ is the output of {Algorithm \ref{alg:glin}}.
\end{theorem}

The above theorem captures the relation between
the first-order approximation error $\varepsilon > 0$ in \eqref{eqn:linear}
and the worst-case approximation ratio of the algorithm.


\section{Faster Batch-Greedy DPP Inference} \label{sec:batch}
In this section, we present an even faster greedy algorithm for the MAP inference task of DPP, in particular
for large-scale tasks.
On top of ideas described in Section \ref{sec:idea1},
we use a batch strategy, i.e., add $k$ elements instead of a single element to the current set,
where $\mathtt{LDAS}$ in Section \ref{sec:naiveimp} is now used as a key component.
The batch strategy accelerates our algorithm.
%
We first provide the formal description of the batch greedy algorithm in Section \ref{sec:algdec2}.
In Section \ref{sec:sharerandom},
we describe additional ideas on applying $\mathtt{LDAS}$ as a subroutine
of the proposed batch algorithm.

\subsection{Algorithm Description} \label{sec:algdec2}

\setcounter{footnote}{1}
\begin{algorithm}[h]
\caption{Faster Batch-Greedy DPP Inference} \label{alg:batch}
\begin{algorithmic}[1]
\STATE {\bf Input:} kernel matrix $L \in \mathbb{R}^{d \times d}$, number of partitions $p$, 
batch size $k$ and the number of batch samples $s$
\STATE {\bf Initialize:} $X \leftarrow \emptyset$
\WHILE {$\mathcal{Y} \setminus X$ is not empty}
\STATE $I_i \leftarrow $ Randomly draw a batch of size $k$ 
for $i \in \left[s\right]$.
\STATE Partition $\left[s\right]$ randomly into $p$ subsets.
\FOR {$j = 1$ { \bfseries to } $p$}
\STATE $\overline{L}_X^{(j)} \leftarrow \text{average of } L_{X \cup I_i}$ for $i$ in the partition $j$
\STATE $Z^{(j)} \leftarrow (|X|+1)$ to $(|X|+k)$-th columns of
$\left( \overline{L}_X^{(j)}\right)^{-1}$
\STATE ${\Gamma}_j \leftarrow \log \det \overline{L}_X^{(j)}$ using $\mathtt{LDAS}$.
\ENDFOR
\FOR{$i = 1$ {\bfseries to } $s$}
\STATE ${\Delta}^{\text{Batch}}_i \leftarrow \left\langle L_{X \cup I_i} - {\overline{L}_{X}^{(j)}}, \mathtt{Mat}\left(Z^{(j)}\right) \right\rangle \footnotemark + \Gamma_j$
\vspace{0.05in}\\
 where a batch index $i$ is included in $j$-th partition.
\ENDFOR
\STATE $i_{\max} \leftarrow \argmax_{i \in [s]} \Delta^{\text{Batch}}_i$ 
\IF{$\log\det L_{X\cup I_{i_{\max}}} - \log\det L_{X} <0$}
\STATE {\bf return $X$}
\ENDIF
\STATE $X \leftarrow X \cup I_{i_{\max}}$
\ENDWHILE
\end{algorithmic}
\end{algorithm}


The formal description of the proposed algorithm is described in {Algorithm \ref{alg:batch}}.
Similar to the line 7 in {Algorithm \ref{alg:glin}}, 
the line 8 of {Algorithm \ref{alg:batch}}
can be solved by the $\mathtt{CG}$ algorithms. 
However, the line 9 of {Algorithm \ref{alg:batch}}
uses the $\mathtt{LDAS}$ 
and we remind that it runs in $\Theta(d^{2})$ time.
In addition, the line 12 requires the vector inner products $k s$ times.
Thus, 
the total complexity becomes
{
$\Theta \left(T_{\mathtt{GR}}^{3}\cdot 
\left( T_{\mathtt{CG}}  + \frac{mn}{k} \right) \cdot p
+ s\cdot T_{\mathtt{GR}}^{2} + s \cdot  T_{\mathtt{CG}}
\right)
= \Theta(T_{\mathtt{GR}}^{3})$
where 
$T_{\mathtt{GR}}$ is the number of greedy updates on the current set $X$
and we choose all parameters $p,T_{\mathtt{CG}},k,s,m,n=O(1)$. 
}
We note that
{Algorithm \ref{alg:batch}} is expected to perform faster than
{Algorithm \ref{alg:glin}} 
when both $T_{\mathtt{GR}}$ and $d$ are large.
This is primarily because
the size of the current set $X$ increases by $k>1$ for each greedy iteration. 
A larger choice of $k$ speeds up the algorithm up to $k$ times, 
but it might hurt its output quality.
We explain more details of key components of the batch algorithm below.

\noindent {\bf Batch selection.} 
The essence of {Algorithm \ref{alg:batch}}
is adding
$k>1$ elements, called batch, simultaneously to the current set with an improved marginal gain.
Formally, it
starts from the empty set and recursively updates $X \leftarrow X \cup I_{\max}$ for
\begin{equation} \label{eqn:batchupdate}
I_{\max} = \argmax_{I \subseteq \mathcal{Y} \setminus X, |I|=k} \log \det L_{X \cup I }.
\end{equation}
until no gain is attained. The non-batch greedy procedure \eqref{eq:maxmarginalgain} corresponds to $k=1$.
Such batch greedy algorithms have been also studied for submodular maximization \cite{nemhauser1978analysis,hausmann1980worst}
and recently, \cite{liu2016performance} studied their theoretical guarantees showing
that they can be better than their non-batch counterparts under some conditions. 
The main drawback of the standard batch greedy algorithms
is that 
finding the optimal batch of size $k$ requires {computing} too many marginal gains of $\binom{|\mathcal{Y} \setminus X|}{k}$ subsets.
To address the issue,
we sample $s \ll \binom{|\mathcal{Y} \setminus X|}{k}$ bunches of batch subsets randomly 
and compute approximate batch marginal gains using them.
\cite{mirzasoleiman2015lazier} first propose 
an uniformly random sampling 
to the standard non-batch greedy {algorithm}.
The authors show that it guarantees $(1-1/e-O(e^{-s}))$ approximation ratio in expectation
and report {that} it performs well in many applications.
In our experiments, we choose $s=50$ batch samples. 

\noindent {\bf High-order approximation of log-determinant.} 
Recall that for {Algorithm \ref{alg:glin}},
we suggest {using the} $\mathtt{CG}$ algorithm under the Schur complement 
for computing 
\begin{equation}\label{eq:avgdiff}
\log \det \overline{L}_{X}^{(j)} - \log \det {L}_{X}.    
\end{equation}
One can apply the same strategy for {Algorithm \ref{alg:batch}},
which requires running the $\mathtt{CG}$ algorithm $k$ times
for \eqref{eq:avgdiff}.
Instead, we suggest {running} $\mathtt{LDAS}$ (using polynomial/high-order approximations
of the scalar log function) only once, i.e., the line 9,
which is much faster if $k$ is large. 
We remind that the asymptotic complexities
of $\mathtt{CG}$ and $\mathtt{LDAS}$ are comparable. 

\subsection{Sharing Randomness in Trace Estimators}\label{sec:sharerandom}
To improve the approximation quality of {Algorithm \ref{alg:batch}},
we further suggest {running} $\mathtt{LDAS}$
using the same random vectors 
$\mathbf{v}^{(1)}, \dots, \mathbf{v}^{(m)}$
across $j \in \left[p\right]$.
This is because we are interested in relative values $\log \det \overline{L}_{X}^{(j)}$ for $j \in \left[p\right]$ 
instead of their absolute ones.

\begin{figure}[h]
\centering
\includegraphics[width=0.6\textwidth]{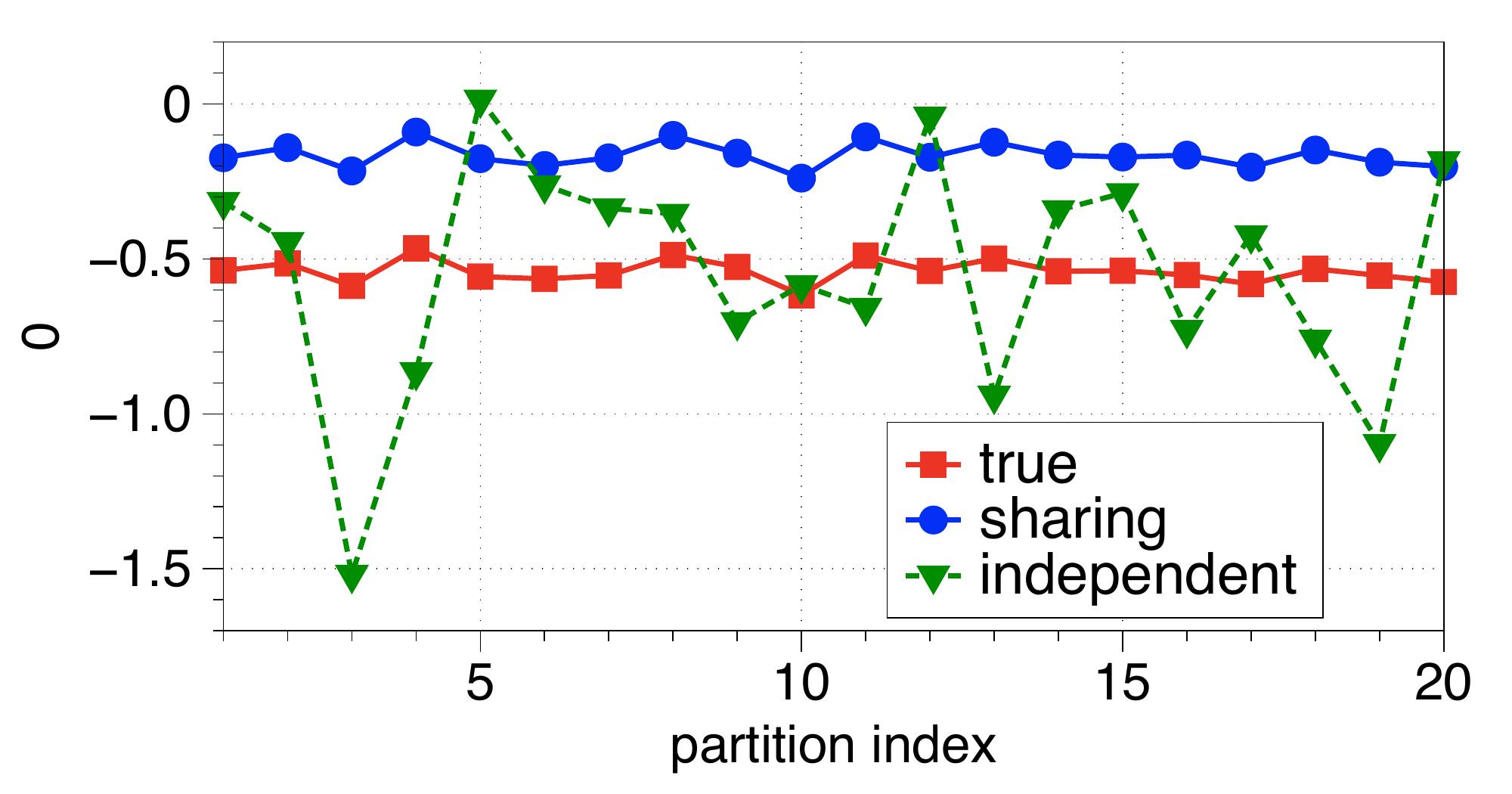} 
\caption{Log-determinant estimation qualities of $\mathtt{LDAS}$ for sharing and independent random vectors.} \label{fig:sharing}
\end{figure}
Our intuition is that 
different random vectors have different bias, which hurt the comparison task. 
Figure \ref{fig:sharing} demonstrates an experiment on the estimation of 
$\log \det \overline{L}_{X}^{(j)}$ when random vectors are 
shared and independent, respectively. 
This implies that
sharing random vectors
might be worse for estimating the absolute values of log-determinants, 
but better for comparing them.

We also formally justify the idea of sharing random vectors as stated in the follows theorem whose
proof is given in Section \ref{sec:pf:thm:sharing}.
\begin{theorem} \label{thm:sharing}
Suppose 
$A,B$ are positive definite matrices 
whose eigenvalues are in $\left[\delta,1-\delta\right]$ for $\delta>0$. 
Let $\Gamma_A, \Gamma_B$ be the estimations of $\log \det A$, $\log \det B$ by $\mathtt{LDAS}$
using {the} same random vectors $\mathbf{v}^{(1)}, \dots, \mathbf{v}^{(m)}$ for both.
Then, it holds that
\begin{align*}
\mathbf{Var}\left[\Gamma_A - \Gamma_B\right] \leq \frac{32M^2 \rho^2 \left(\rho+1\right)^2}{m\left(\rho-1\right)^6 \left( 1-2\delta\right)^2} \| A-B \|_{F}^2
\end{align*}
where $M = 5\log\left(2/\delta\right)$ and $\rho=1 + \frac{2}{\sqrt{2/\delta - 1}-1}$.
\end{theorem}

Without sharing random vectors, 
the variance 
should grow linearly with respect to $\norm{A}_F^2+\norm{B}_F^2$.
In our case, matrices $A$ and $B$ correspond to some of $\overline{L}_{X}^{(j)}$,
and $\| A-B \|_{F}^2$ {is} significantly smaller than $\norm{A}_F^2+\norm{B}_F^2$.
We believe that our idea of sharing randomness might be of broader interest
in many applications of $\mathtt{LDAS}$ or its variants, requiring
multiple log-determinant computations.
\section{Proof of Theorems}\label{sec:proof}
In this section, 
we provide the proof of our main theorems.

\subsection{Proof of Theorem \ref{thm:alg1}}\label{sec:pf:thm:alg1}
For given $X \subseteq \mathcal{Y}$,
we denote that 
the true marginal gain $\Lambda_i$
and the approximated gain $\Delta_i$ (used in {Algorithm \ref{alg:glin}}) as
\begin{align*}
\Lambda_i &:= \log \det L_{X \cup \{ i\}} - \log \det L_X,\\
\Delta_i &:= 
\left\langle {\left(\overline{L}_{X}^{(j)}\right)^{-1}}, L_{X \cup \{ i \}} 
- {\overline{L}_{X}^{(j)}} \right\rangle 
+ \left(\log \det \overline{L}_{X}^{(j)} - \log \det {L}_{X}\right)
\end{align*}
where an item $i\in \mathcal{Y}\setminus X$ is in the partition $j$.
We also use $i_{\mathtt{OPT}} = \argmax_i \Lambda_i$ and 
$i_{\max} = \argmax_i \Delta_i$.
Then, we have 
\begin{align*}
\Lambda_{i_{\max}}
{\geq} \ \Delta_{i_{\max}} - \varepsilon \
{\geq} \ \Delta_{i_{\mathtt{OPT}}} - \varepsilon \
{\geq} \ \Lambda_{i_{\mathtt{OPT}}} - 2 \varepsilon \
\end{align*}
where the first and third inequalities are 
from the definition of $\varepsilon$, i.e., $\abs{\Lambda_i - \Delta_i } \leq \varepsilon$,
and the second inequality holds by the optimality of $i_{\max}$.
In addition,
when the smallest eigenvalue of $L$ is greater than 1, 
$\log \det L_X$ is monotone and non-negative \cite{sharma2015greedy}.
To complete the proof, 
we introduce
following approximation guarantee 
of the greedy algorithm with a `noise' during the selection \cite{streeter2009online}.

\begin{theorem}{\bf (Noisy greedy algorithm)}\label{thm:egreedy}
Suppose a submodular function $f$ defined on ground set $\mathcal{Y}$
is monotone and non-negative.
Let $X_0 = \emptyset$ and 
$X_k = X_{k-1} \cup \{ i_{\max} \}$
such that 
\begin{align*}
&f(X_{k-1} \cup \{ i_{\max}\}) - f(X_{k-1}) 
\geq \max_{i \in \mathcal{Y} \setminus X_{k-1}} \left( f(X_{k-1} \cup \{ i\}) - f(X_{k-1})\right) - \varepsilon_k
\end{align*}
for some $\varepsilon_k \geq 0$.
Then, 
$$
f(X_k) \geq \left(1- 1/e \right) \max_{X \subseteq \mathcal{Y}, |X|\leq k} f(X) 
- \sum_{i=1}^k \varepsilon_i
$$
\end{theorem}

Theorem \ref{thm:alg1} is straightforward by
substituting $2 \varepsilon$ into $\varepsilon_k$.
This completes the proof of Theorem \ref{thm:alg1}.


\subsection{Proof of Theorem \ref{thm:sharing}}\label{sec:pf:thm:sharing}
As we explained in Section \ref{sec:naiveimp},
Chebyshev expansion of $\log x$ in $[\delta, 1-\delta]$ with degree $n$ 
is defined as $\pn{x}$. This can be written as
\begin{align} 
\pn{x} = \sum_{k=0}^n c_k T_k\left(\frac{2}{1-2\delta}x - \frac1{1-2\delta}\right) \label{eqn:pn}
\end{align}
where the coefficient $c_k$ and the $k$-th Chebyshev polynomial $T_k(x)$ are defined as
\begin{align}
&c_k
= \begin{dcases}
      \frac{1}{n+1}  \sum_{j=0}^n f\left(\frac{1-2\delta}{2}x_j+\frac12\right) \ T_0(x_j) & \text{if $\ k=0$} \\
      \frac{2}{n+1}  \sum_{j=0}^n f\left(\frac{1-2\delta}{2}x_j+\frac12\right) \ T_k(x_j) & \text{otherwise}
      \end{dcases} \label{eqn:ck} \\
&T_{k+1}(x) = 2 x T_k (x) - T_{k-1} (x) \label{eqn:recur} \qquad \text{for $\ k \ge 1$} 
\end{align}
where $x_j = \cos \left( \frac{ \pi (j + 1/2 )}{ n+1} \right)$ for $j = 0,1,\dots,n$ and $T_0(x) = 1$, $T_1(x) = x$ \cite{mason2002chebyshev}.
For simplicity, we now use $H := \pn{A} - \pn{B}$
and denote $\widetilde A = \frac{2}{1-2\delta}A - \frac1{1-2\delta} \mathbf{I}$ 
where $\mathbf{I}$ is identity matrix with same dimension of $A$ and
same for $\widetilde B$.

We estimate the log-determinant difference while random vectors are shared, i.e.,
\begin{align*}
\log \det A - \log \det B 
\approx 
\frac1m \sum_{i=1}^m \mathbf{v}^{(i)\top} H\mathbf{v}^{(i)}.
\end{align*}
To show that the variance of $\mathbf{v}^{(i)\top} H \mathbf{v}^{(i)}$ is small as $\norm{A - B}_F$,
we provide that
\begin{align*}
\mathbf{Var}\left[\frac1m \sum_{i=1}^m \mathbf{v}^{(i)\top} H\mathbf{v}^{(i)}\right] 
&= \frac1m \mathbf{Var}\left[\mathbf{v}^\top H \mathbf{v}\right] \\
&\leq \frac2m \norm{H}_F^2 = \frac2m \norm{\pn{A}-\pn{B}}_F^2 \\
&\leq \frac2m \left( \sum_{k=0}^n \abs{c_k} \norm{T_{k} \left(\widetilde A\right) - T_{k} \left(\widetilde B\right)}_F\right)^2
\end{align*}
where the first inequality holds from \cite{avron2011randomized}
and the second is from combining \eqref{eqn:pn} with the triangle inequality.
To complete the proof, we use the following two lemmas.

\begin{lemma} \label{lmm:chebpolbound}
Let $T_k \left( \cdot \right)$ be Chebyshev polynomial with $k$-degree 
and  symmetric matrices $B, E$ 
satisfied with $\norm{B}_2 \leq 1$, $\norm{B + E}_2 \leq 1$.
Then, for $k \geq 0$,
\begin{align*}
\norm{T_k\left( B + E\right) - T_k\left( B\right)}_F \leq k^2 \norm{E}_F.
\end{align*}
\end{lemma}

\begin{lemma} \label{lmm:coeffbound}
Let $c_k$ be the $k$-th coefficient of Chebyshev expansion for $f\left(x\right)$.
Suppose $f$ is analytic with $\abs{f\left(z\right)}\leq M$ in the region bounded by the ellipse with foci $\pm 1$
and the length of major and minor semiaxis summing to $\rho>1$. Then,
\begin{align*}
\sum_{k=0}^n k^2 \abs{c_k} \leq \frac{2M \rho\left(\rho+1\right)}{\left(\rho-1\right)^3}.
\end{align*}
\end{lemma}

In order to apply Lemma \ref{lmm:coeffbound},
we should consider $f(x) = \log \left( \frac{1-2\delta}{2} x +\frac12 \right)$.
Then it can be easily obtained $M = 5 \log\left(2/\delta\right)$ and $\rho =1 + \frac{2}{\sqrt{2/\delta - 1}-1}$ as provided in \cite{han2015large}.

Using Lemma \ref{lmm:chebpolbound} and \ref{lmm:coeffbound}, 
we can write
\begin{align*}
\mathbf{Var}\left[\frac1m \sum_{i=1}^m \mathbf{v}^{(i)\top} H\mathbf{v}^{(i)}\right]  
&\leq \frac2m \left( \sum_{k=0}^n \abs{c_k} \norm{T_{k} \left(\widetilde A\right) - T_{k} \left(\widetilde B\right)}_F\right)^2 \\
&\leq \frac2m \left( \sum_{k=0}^n \abs{c_k} k^2 \norm{\widetilde A - \widetilde B}_F\right)^2 \\
&\leq \frac2m \left( \frac{2M \rho \left(\rho+1 \right)}{\left(\rho -1\right)^3}\right)^2 \left( \frac{2}{1-2\delta}\norm{A-B}_F\right)^2 \\
&= \frac{32M^2\rho^2\left( \rho+1\right)^2}{m\left(\rho-1\right)^6\left(1-2\delta\right)^2} \norm{A- B}_F^2
\end{align*}
where the second inequality holds from Lemma \ref{lmm:chebpolbound}
and the thrid is from Lemma \ref{lmm:coeffbound}.
This completes the proof of Theorem \ref{thm:sharing}.

\subsection{Proof of Lemma \ref{lmm:chebpolbound}}
Denote $R_k := T_{k} \left(B + E\right) - T_{k} \left( B\right)$.
From the recurrence of Chebyshev polynomial \eqref{eqn:recur},
$R_k$ has following 
\begin{align} 
&R_{k+1}
= 2\left( B + E\right) R_k - R_{k-1} + 2 E \ T_k\left( B\right) \label{eqn:recurR}
\end{align}
for $k\geq 1$ where $R_1 = E$, $R_0 = \mathbf{0}$
where $\mathbf{0}$ is defined as zero matrix with the same dimension of $B$.
Solving this, we obtain that 
\begin{align}
R_{k+1} = g_{k+1}\left( B + E\right) E 
+ \sum_{i=0}^{k} h_i \left(B + E\right)  E \ T_{k+1-i} \left(B\right) \label{eq:closedR}
\end{align}
for $k\geq 1$ where both $g_k\left(\cdot\right)$ and $h_k\left(\cdot\right)$ are polynomials 
with degree $k$ 
and they have following recurrences 
\begin{align*}
g_{k+1}\left(x\right) &= 2 x g_{k}\left(x\right) - g_{k-1}\left(x\right), g_1\left(x\right) = 1, g_0\left( x\right) = 0, \\
h_{k+1}\left(x\right) &= 2 x h_{k}\left(x\right) - h_{k-1}\left(x\right), h_1\left(x\right) = 2, h_0\left( x\right) = 0.
\end{align*}
In addition, we can easily verify that 
$$2 \max_{x \in \left[-1,1\right]} \abs{g_k \left(x\right)} = \max_{x \in \left[-1,1\right]} \abs{h_k \left(x\right)} = 2k.$$
Putting all together, we conclude that
\begin{align*}
\norm{R_{k+1}}_F 
&\leq \norm{g_{k+1}\left( B + E\right) E }_F  
+ \norm{\sum_{i=0}^{k} h_i \left(B + E\right)  E \ T_{k+1-i} \left(B\right)}_F \\
&\leq \norm{g_{k+1}\left( B + E\right)}_2 \norm{E }_F 
 + \sum_{i=0}^{k} \norm{h_i \left(B + E\right)}_2 \norm{E}_F \norm{\ T_{k+1-i} \left(B\right)}_2\\
&\leq \left( \norm{g_{k+1}\left( B + E\right)}_2 + \sum_{i=0}^{k} \norm{h_i \left(B + E\right)}_2\right) \norm{E}_F \\
&\leq \left( k+1 + \sum_{i=0}^k 2 i\right) \norm{E}_F \\
&= \left(k+1\right)^2 \norm{E}_F
\end{align*}
where the second inequality holds from $\norm{YX}_F = \norm{XY}_F \leq \norm{X}_2 \norm{Y}_F$
for matrix $X,Y$
and the third inequality uses that $\abs{T_k\left(x\right)} \leq 1$ for all $k\geq 0$.
This completes the proof of Lemma \ref{lmm:chebpolbound}.

\subsection{Proof of Lemma \ref{lmm:coeffbound}}
For general analytic function $f$, 
Chebyshev series of $f$ is defined as 
\begin{align*}
f\left(x\right) = \frac{a_0}2 + \sum_{k=1}^\infty a_k T_k\left(x\right), \quad a_k = \frac2\pi \int_{-1}^1 \frac{f\left(x\right) T_k\left(x\right)}{\sqrt{1-x^2}} dx.
\end{align*}
and from \cite{mason2002chebyshev} it is known that 
\begin{align*}
c_k - a_k = \sum_{j=1}^{\infty} \left(-1\right)^j \left(a_{2j(n+1)-k} + a_{2j(n+1)+k}\right)
\end{align*}
and 
$
\abs{a_k} \leq \frac{2M}{\rho^k}
$
for $0 \leq k \leq n$. We remind that $c_k$ is defined in \eqref{eqn:ck}.
Using this facts, we get
\begin{align*}
k^2 \abs{c_k} 
&\leq k^2 \left( \abs{a_k} + \sum_{j=1}^\infty \abs{a_{2j(n+1)-k}} + \abs{a_{2j(n+1)+k}}\right) \\
&\leq k^2 \abs{a_k} + \sum_{j=1}^\infty k^2 \abs{a_{2j(n+1)-k}} + k^2 \abs{a_{2j(n+1)+k}}\\
&\leq k^2 \abs{a_k} + \sum_{j=1}^\infty \left(2j(n+1)-k\right)^2 \abs{a_{2j(n+1)-k}}  
 + \left(2j(n+1)+k\right)^2 \abs{a_{2j(n+1)+k}}
\end{align*}
Therefore, we have
\begin{align*}
\sum_{k=0}^n k^2 \abs{c_k} 
\leq \sum_{k=0}^n k^2 \abs{a_k}
\leq \sum_{k=0}^\infty k^2 \abs{a_k} \leq \sum_{k=0}^\infty k^2 \frac{2M}{\rho^k} = \frac{2M \rho\left(\rho+1\right)}{\left(\rho-1\right)^3}
\end{align*}
This completes the proof of Lemma \ref{lmm:coeffbound}.

\section{Experimental Results} \label{sec:exp}
In this section, we evaluate our proposed algorithms for the MAP inference on synthetic
and real-world DPP instances. 
\footnote{The codes are available in \url{https://github.com/insuhan/fastdppmap}.}

\noindent {\bf Setups.}
The experiments are performed 
using a machine with {a hexa-core} Intel CPU (Core i7-5930K, 3.5 GHz) and 32 GB RAM. 
We compare our algorithms with following competitors:
the lazy greedy algorithm (\textsc{Lazy}) \cite{minoux1978accelerated},
double greedy algorithm (\textsc{Double}) \cite{buchbinder2015tight} and
softmax extension (\textsc{Softmax}) \cite{gillenwater2012near}.
In all our experiments, 
\textsc{Lazy} is significantly faster than 
the na\"ive greedy algorithms 
described in Section \ref{sec:naiveimp}, 
while they produce the same outputs.
Hence, we
use \textsc{Lazy} as the baseline of evaluation.
Unless stated otherwise, 
we choose parameters of $p=5$, $k=10$, $s=50$, $m=20$ and $n=15$,
regardless matrix dimension, for our algorithms. 
{
We also run $\mathtt{CG}$ until it achieves convergence error less than $10^{-10}$
and typically $T_{\mathtt{CG}} \leq 30$.
}

\noindent {\bf Additional tricks for boosting accuracy.} 
For boosting approximation qualities of our algorithms,
we use the simple trick in our experiments: recompute
top $\ell$ marginal gains exactly (using $\mathtt{CG}$) where
they are selected based on estimated marginal gains, i.e.,
$\Delta_i$ for {Algorithm \ref{alg:glin}}
and $\Delta^{\text{Batch}}_i$ for {Algorithm \ref{alg:batch}}.
Then, our algorithms choose the best element among $\ell$ candidates,
based on their exact marginal gains.
Since we choose small $\ell=20$ in our experiments, 
this additional process increases the running times of our algorithms marginally,
but makes them more accurate.
In fact, the trick is inspired from \cite{minoux1978accelerated}
where the authors also recompute the exact marginal gains of few elements.
In addition, for boosting further approximation qualities of
{Algorithm \ref{alg:batch}}, we also run {Algorithm \ref{alg:glin}}
in parallel and choose the largest one among $\{\Delta_i, 
\Delta^{\text{Batch}}_i\}$ given the current set. 
Hence, at most iterations, the batch with the maximal $\Delta^{\text{Batch}}_i$ is chosen
and increases the current set size by $k$ (i.e., making speed-up) as like {Algorithm \ref{alg:batch}},
and the non-batch with the maximal $\Delta_i$ is chosen at very last iterations, 
which fine-tunes the solution quality.
We still call the synthesized algorithm by {Algorithm \ref{alg:batch}} in this section.

\noindent {\bf Performance metrics.}
For the performance measure on approximation qualities of algorithms,
we use the following ratio of log-probabilities:
$$
{\log \det L_X}/{\log \det L_{X_{\textsc{Lazy}}}}.
$$
where 
$X$ and $X_{\textsc{Lazy}}$ are the outputs of an algorithm and 
\textsc{Lazy}, respectively. Namely, we compare outputs of algorithms with that of
\textsc{Lazy} since the exact optimum is hard to compute.
Similarly, we report the running time speedup of each algorithm over \textsc{Lazy}.

\begin{figure*}[ht]
\begin{center}
\subfigure[]{\includegraphics[width=0.4\textwidth]{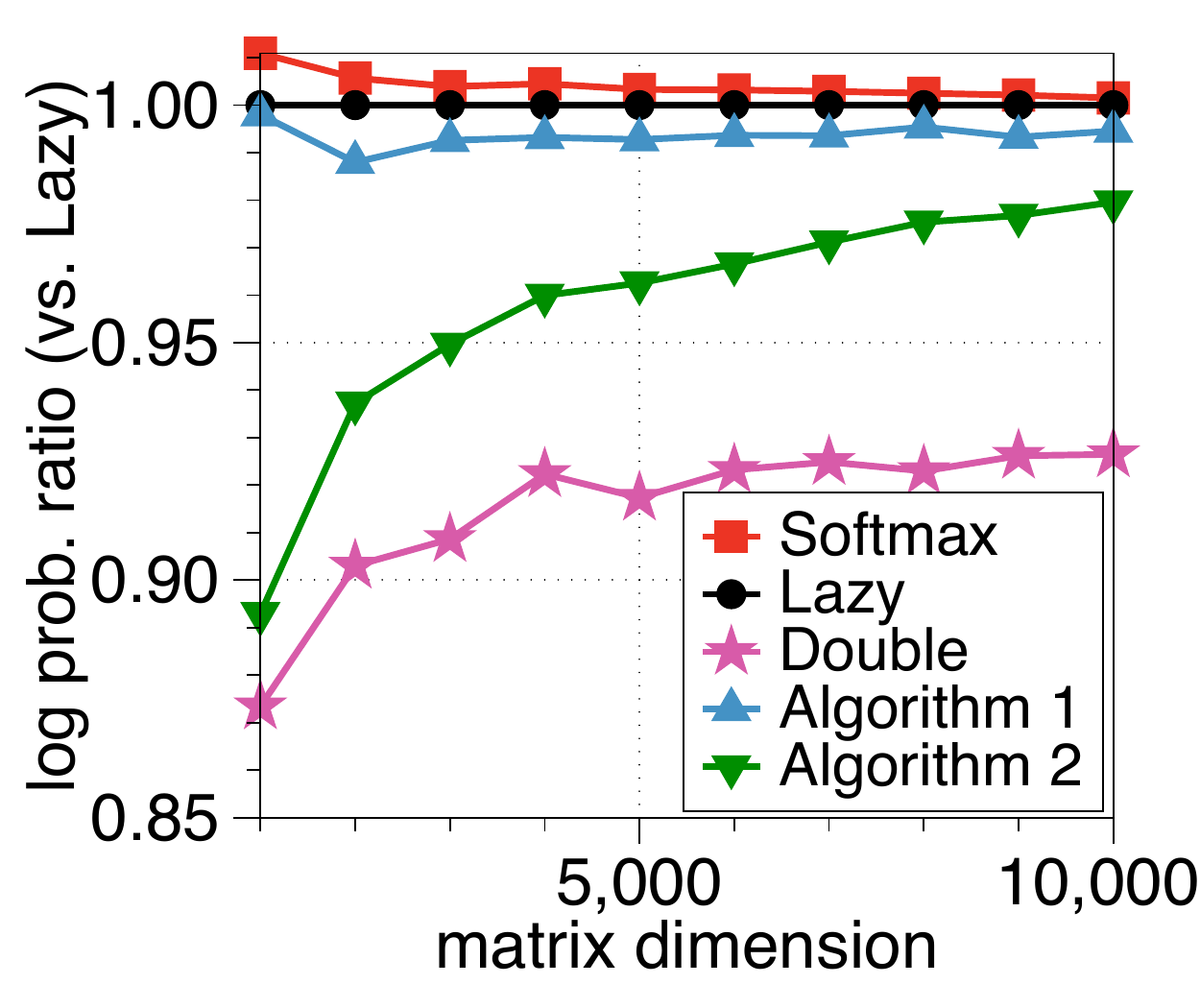}}
\label{fig:syn:logp}
\hspace{-0.12in}
\subfigure[]{\includegraphics[width=0.4\textwidth]{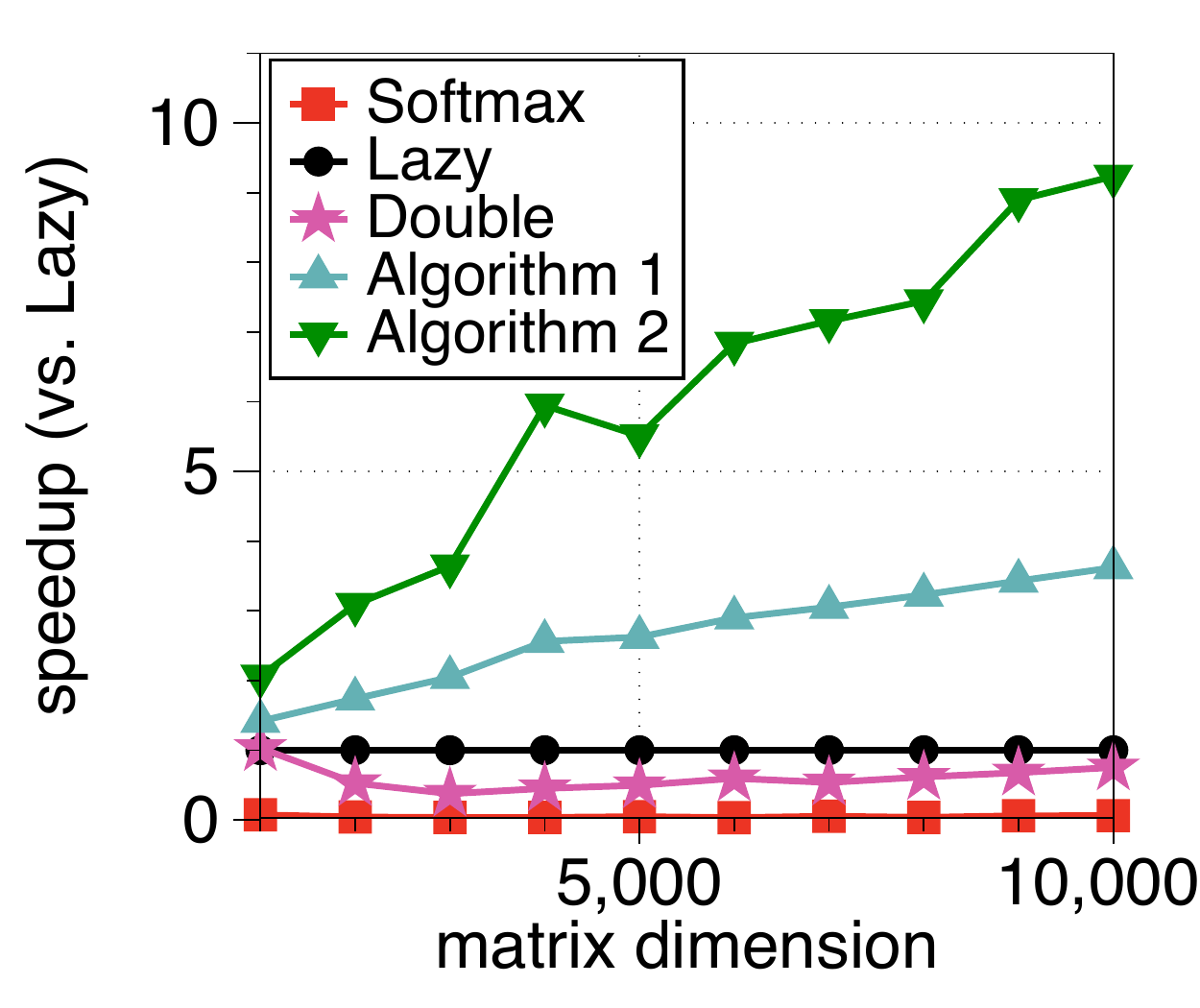}}
\hspace{-0.12in}
\vskip -0.15in
\subfigure[]{\includegraphics[width=0.4\textwidth]{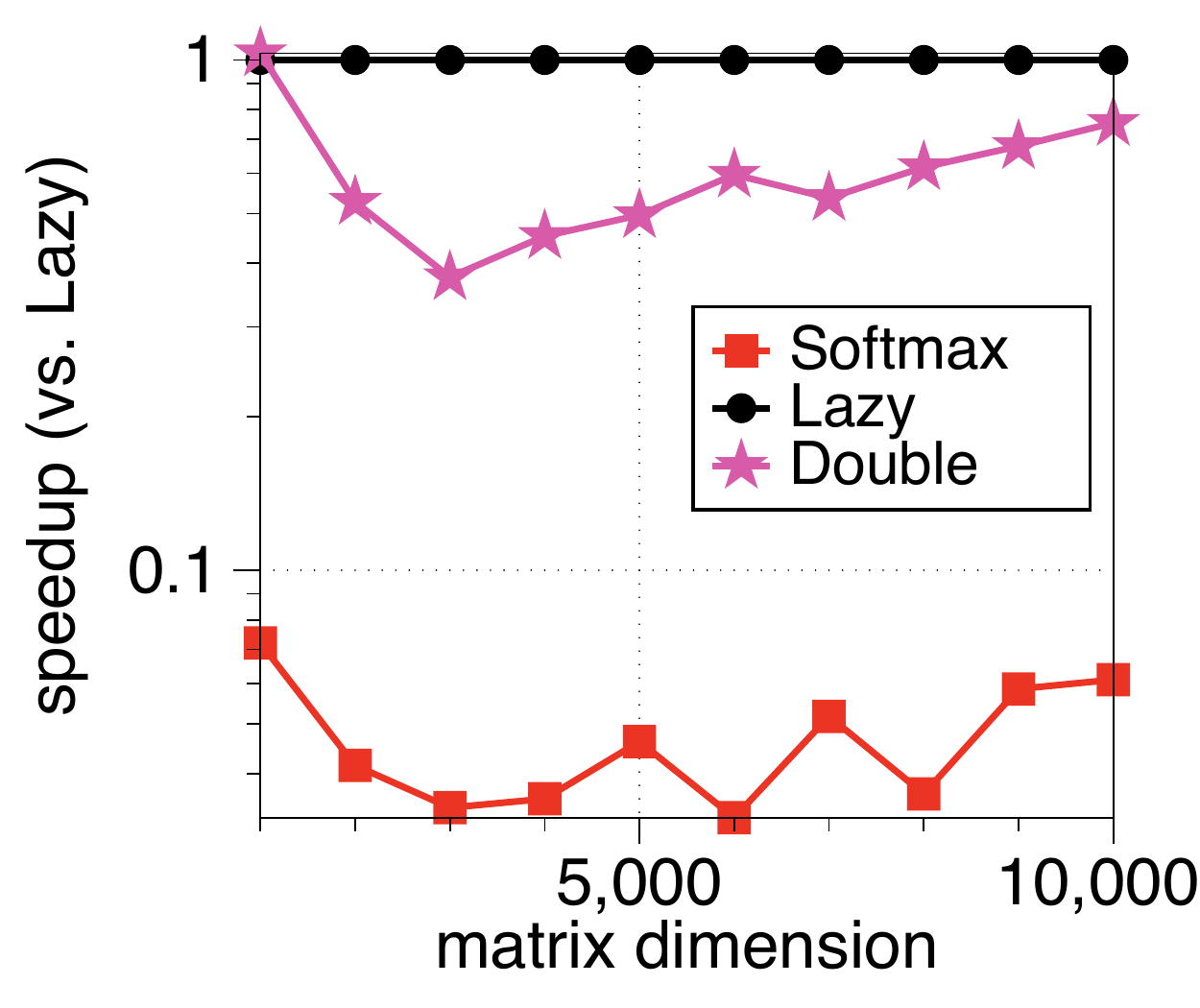}}
\hspace{-0.12in}
\subfigure[]{\includegraphics[width=0.4\textwidth]{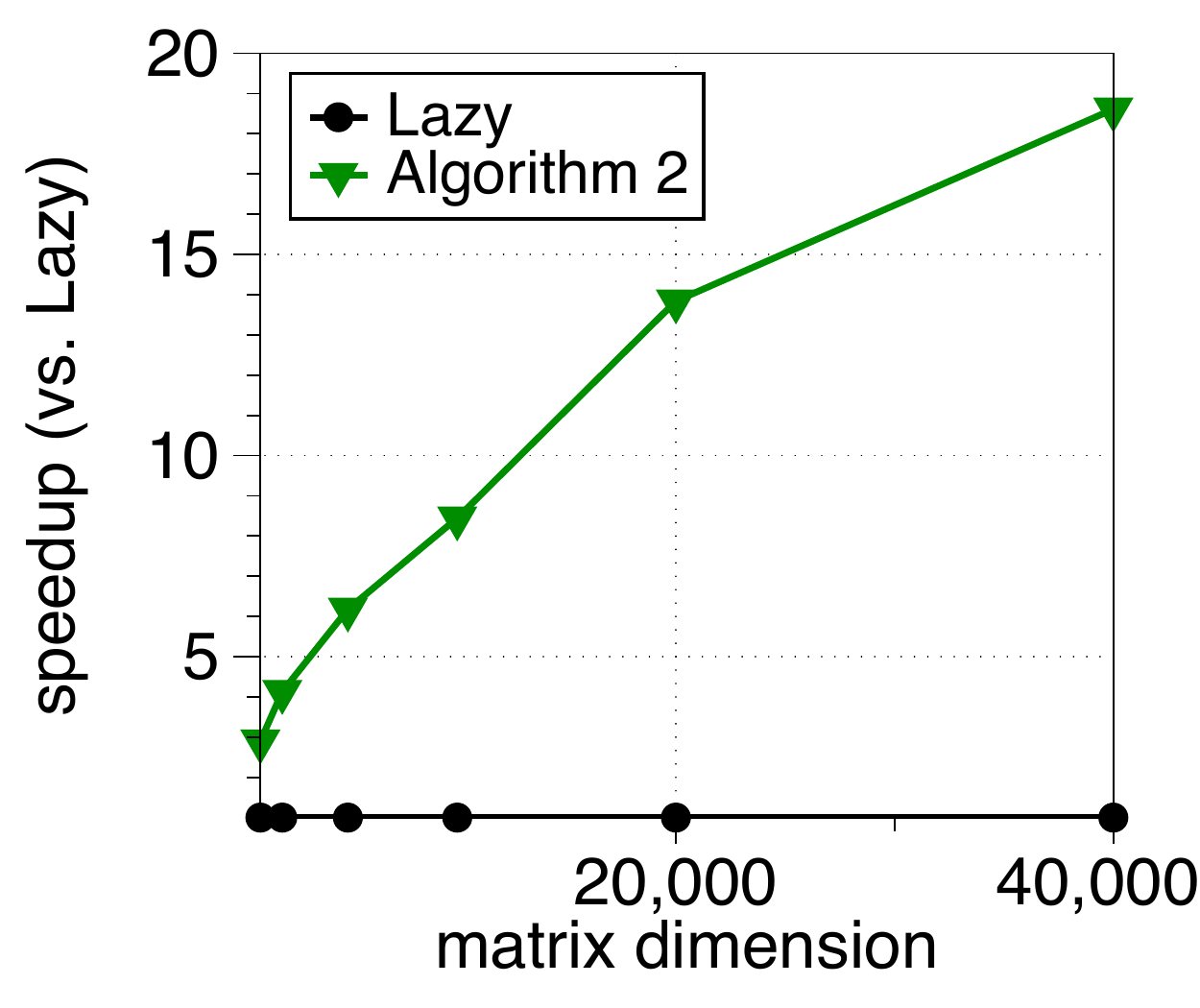}}
\label{fig:syn:speed}
\vskip -0.15in
\caption{Plot of (a) log-probability ratio and (b), (c) and (d) implies speedup for \textsc{Softmax}, \textsc{Double}, {Algorithm \ref{alg:glin}} and {Algorithm \ref{alg:batch}} compared to \textsc{Lazy}, respectively. 
{Algorithm \ref{alg:glin}} is about $3$ times faster the lazy greedy algorithm (\textsc{Lazy})
while loosing only $0.2\%$ accuracy at $d=10,000$.
{Algorithm \ref{alg:batch}} has $2 \%$ loss on accuracy 
but $9$ times faster than \textsc{Lazy} at $d=10,000$.
If dimension is $d=40,000$, it runs $19$ times faster.}
\label{fig:syn}
\end{center}
\end{figure*} 

\subsection{Synthetic Dataset}

In this section, we use synthetic DPP datasets generated as follows.
As \cite{kulesza2011learning,kulesza2012determinantal} proposed,
a kernel matrix $L$ for DPP can be re-parameterized as 
$$L_{i,j} = q_i {\phi}_i^\top {\phi}_j q_j,$$
where $q_i \in \mathbb{R}^{+}$ is considered as the quality of item $i$
and $\mathbf{\phi}_i \in \mathbb{R}^{d}$ is the normalized feature vector of  item $i$
so that $\mathbf{\phi}_i^\top \mathbf{\phi}_j$ measures the similarity between $i$ and $j$.
We use $q_i = \exp\left( \beta_1 {x}_i + \beta_2 \right)$ for the quality measurement ${x}_i \in \mathbb{R}$ and choose $\beta_1 = 0.01, \beta_2 = 0.2$. 
We choose each entry of $\phi_i$ and ${x}_i$ drawn from the normal distribution
$\mathcal{N}(0,1)$ for all $i \in \left[ d\right]$, and
then normalize $\phi_i$ so that $\norm{\phi_i}_2 = 1$.

\begin{figure}[t]
\begin{center}
\subfigure[]{\includegraphics[width=0.4\textwidth]{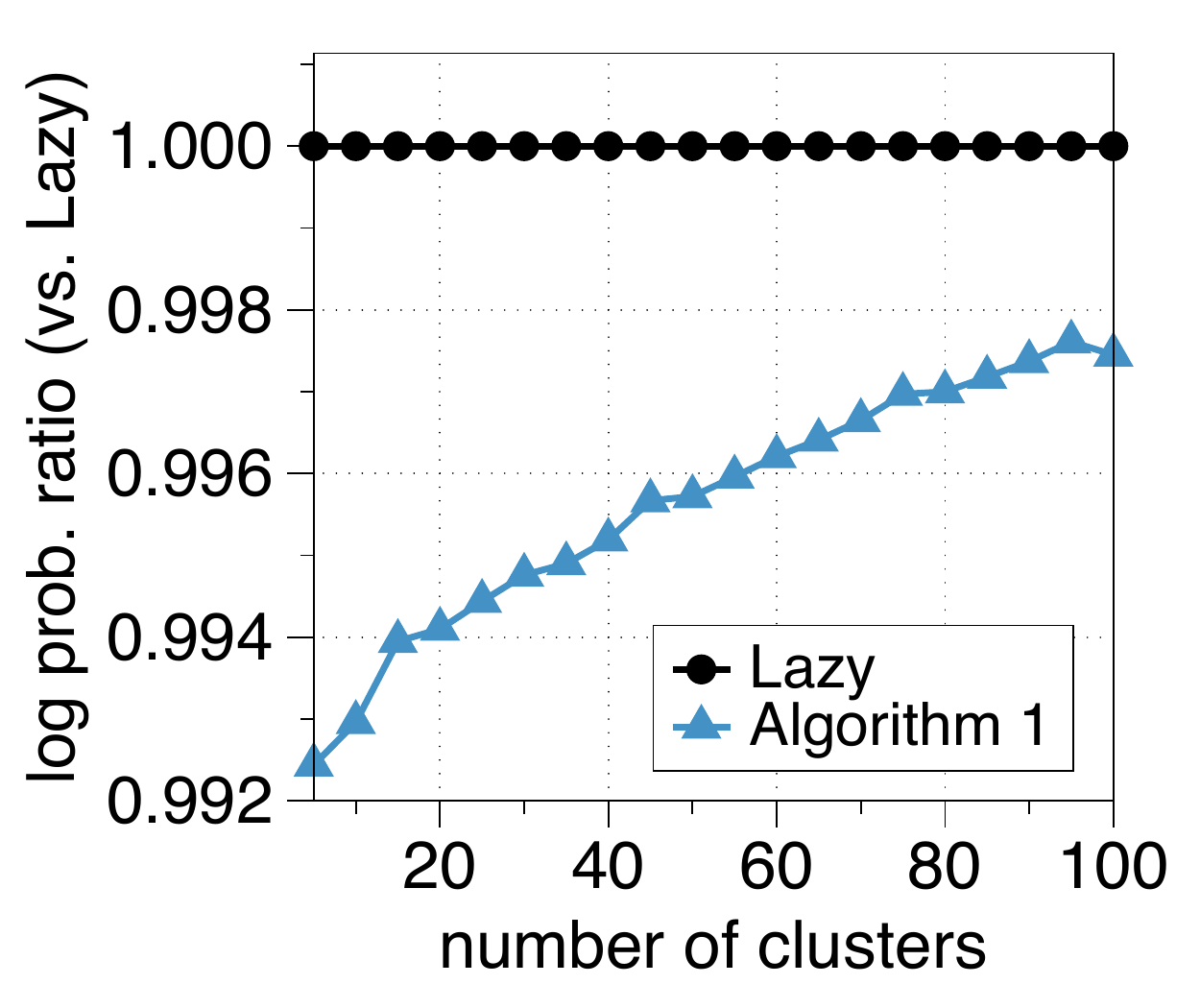}\label{fig:cluster}} 
\subfigure[]{\includegraphics[width=0.4\textwidth]{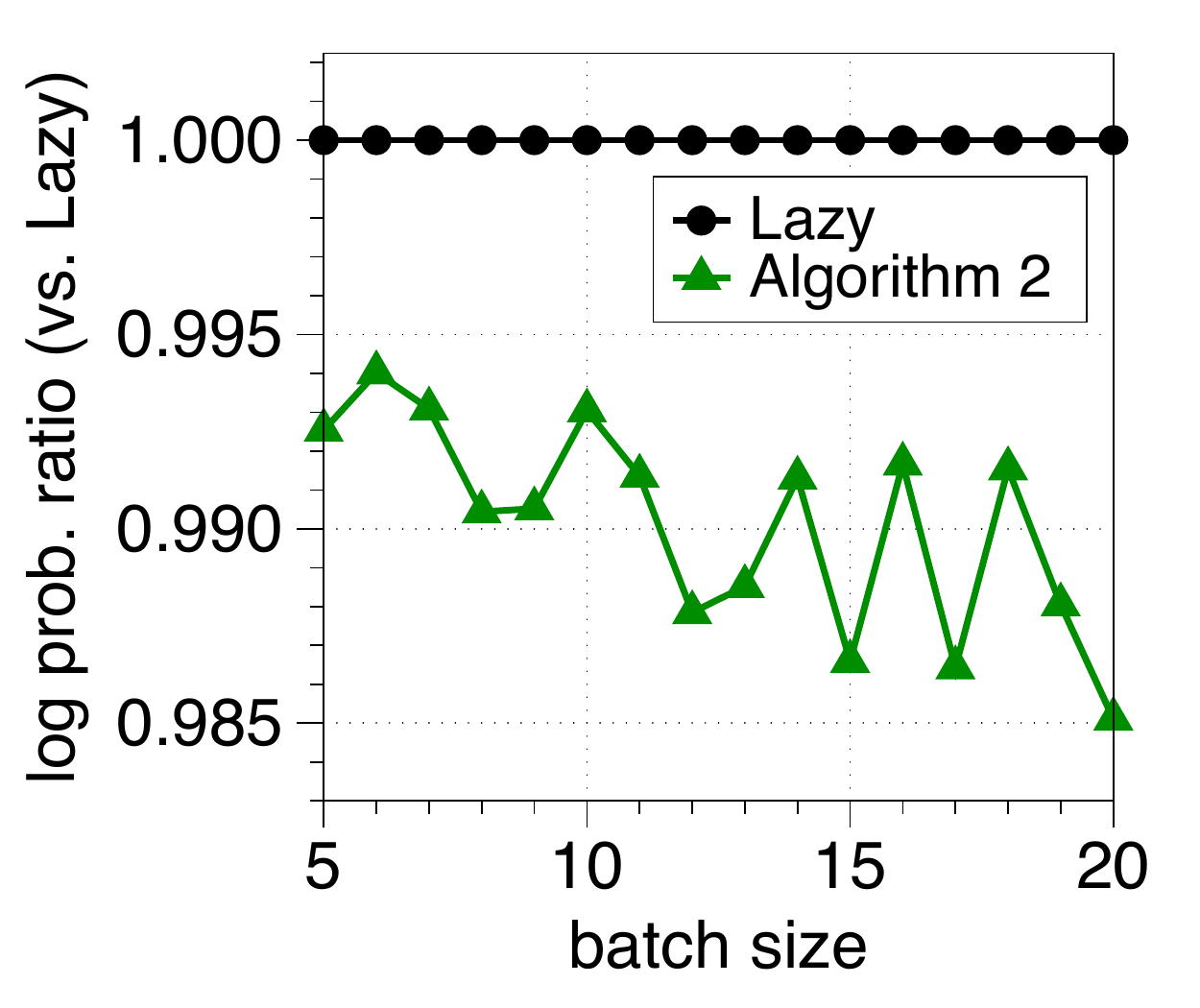}\label{fig:batch}}
\vskip -0.15in
\caption{Log-probability ratios compared to \textsc{Lazy}:
(a) {Algorithm \ref{alg:glin}} changing the number of clusters $p$
and (b) 
{Algorithm \ref{alg:batch}} varying the batch size $k$. These experiments are done under {$d=1,000$}.}
\label{fig:perform}
\end{center}
\end{figure}

We first show how much the number of clusters $p$ and the batch size $k$ are sensitive for
{Algorithm \ref{alg:glin}} and {Algorithm \ref{alg:batch}}, respectively.
Figure \ref{fig:cluster} shows the accuracy of {Algorithm \ref{alg:glin}} with different numbers of clusters.
It indeed confirms that a larger cluster improves its accuracy since
it makes first-order approximations tighter.
Figure \ref{fig:batch} shows the performance trend of {Algorithm \ref{alg:batch}} as the batch size $k$ increases, which shows that 
a larger batch might hurt its accuracy.
Based on these experiments, we choose $p=5, k=10$ in order to target
$0.01$ approximation ratio loss compared to \textsc{Lazy}.

We generate synthetic kernel matrices with varying dimension $d$ up to $40,000$,
and the performances of tested algorithms are reported in Figure \ref{fig:syn}(a).
One can observe that
\textsc{Lazy} seems to be near-optimal, where
only \textsc{Softmax} often provides marginally larger log-probabilities 
than \textsc{Lazy} under small dimensions.
{
Interestingly, we found that 
\textsc{Double} has the strong theoretical guarantee for general submodular 
maximization \cite{buchbinder2015tight},
but its practical performance for DPP is worst among evaluating algorithms. 
Moverover, it is slightly slower than \textsc{Lazy}. 
In summary, one can conclude that 
our algorithms can be at orders of magnitude faster 
than \textsc{Lazy}, \textsc{Double} and \textsc{Softmax}, 
while loosing $0.01$-approximation ratio.
For example, {Algorithm \ref{alg:batch}} is 
$19$ times faster than \textsc{Lazy} for $d=40,000$,
and the gap should increase for larger dimension $d$.
}

\subsection{Real Dataset}
We use real-world datasets of the following two tasks of
matched and video summarizations.

\begin{figure}[h]
\begin{center}
\subfigure[]{\includegraphics[width=0.4\textwidth]{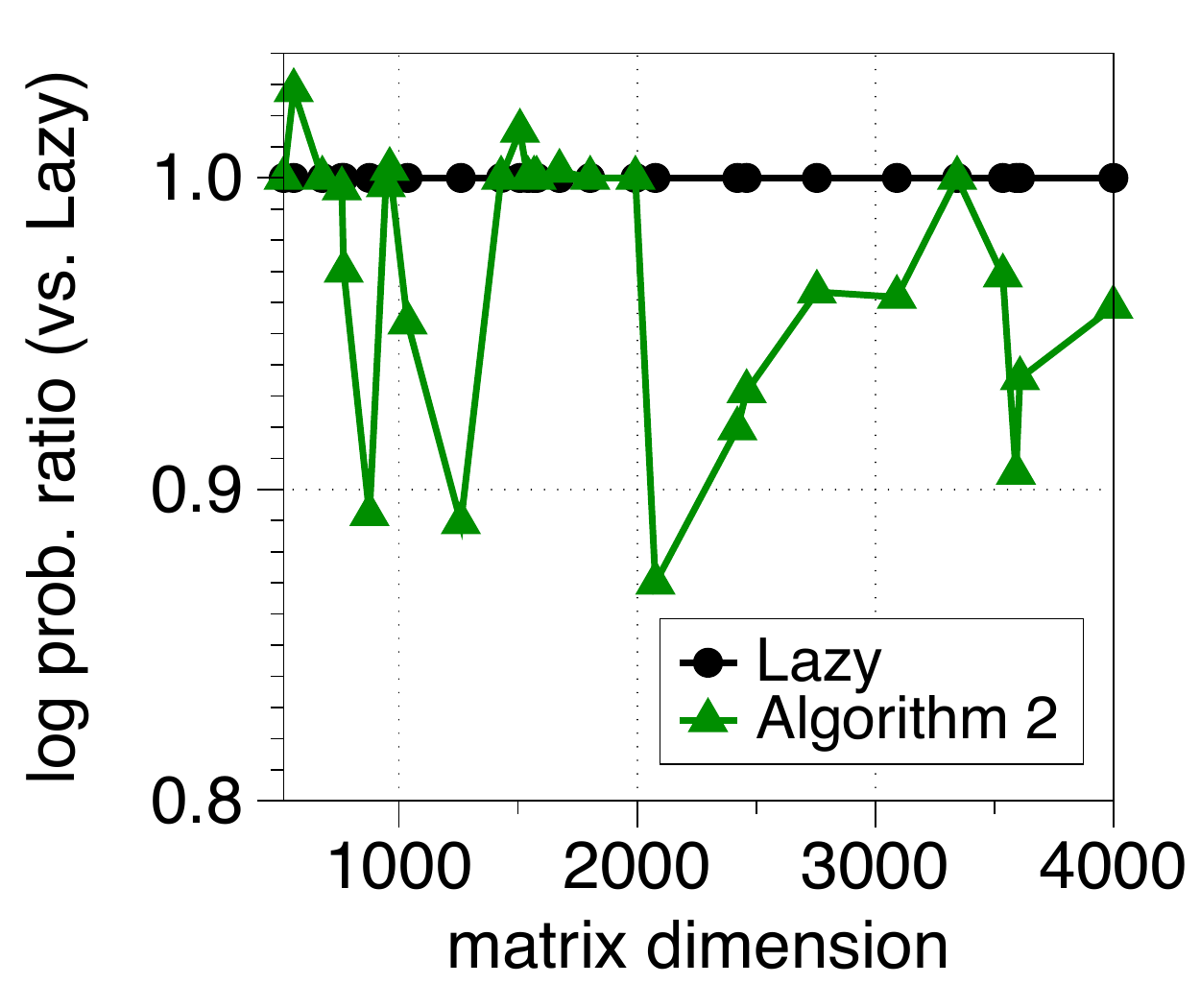} \label{fig:match:logp}}
\subfigure[]{\includegraphics[width=0.4\textwidth]{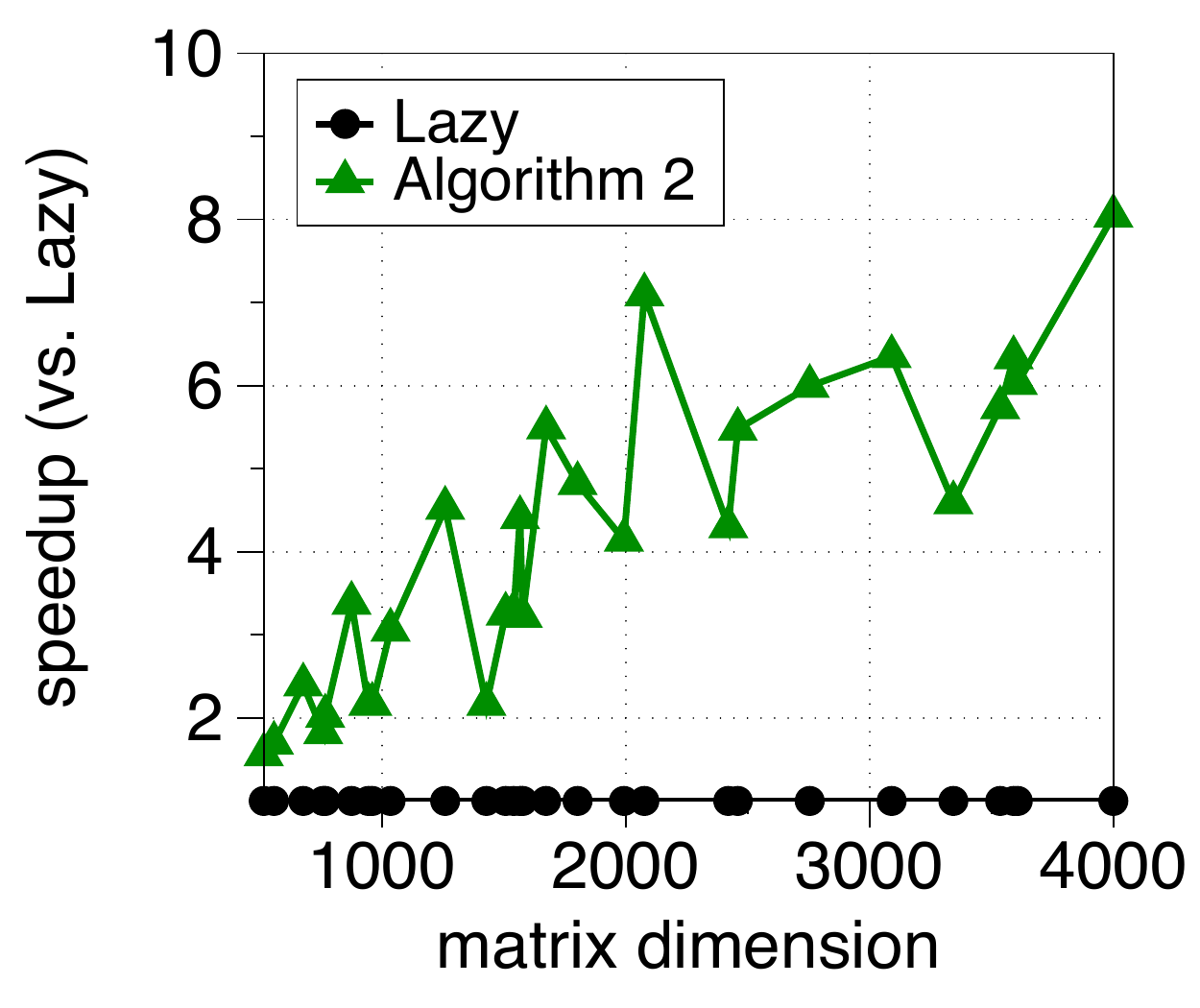}\label{fig:match:speedup}}
\vskip -0.15in
\caption{Plot of log-probability ratio and speedup (log-scale) of {Algorithm \ref{alg:batch}}, compared to  \textsc{Lazy}, for matched summarization under 2016 Republican presidential primaries.}
\label{fig:match}
\end{center}
\end{figure}

{\bf Matched summarization.}
We evaluate our proposed algorithms 
for matched summarization {that} is first proposed by \cite{gillenwater2012near}.
This task gives useful information for comparing {the} texts 
addressed at different times by the same speaker.
Suppose we have two different documents 
and each one consists of several statements.
The goal is to apply DPP for finding statement pairs 
{that} are similar to each other, while they
summarize (i.e., diverse) well the two documents. 
We use transcripts of debates in 2016 US Republican party presidential primaries
speeched by following $8$ participates:
Bush, Carson, Christie, Kasich, Paul, Trump, 
Cruz and Rubio.\footnote{Details of the primaries are provided in
 \url{http://www.presidency.ucsb.edu/debates.php}.}

We follow similar pre-processing steps 
of \cite{gillenwater2012near}.
First, every sentence is parsed and only nouns except the stopwords {are extracted} via NLTK \cite{bird2006nltk}.
Then, {we remove the} `rare' words occurring less than $10 \%$ of the whole debates,
and then ignore each statement which contains more `rare' words than 'frequent' ones in it.
This gives us a dataset containing $3,406$ distinct `frequent' words and $1,157$ statements.
For each statement pair $(i,j)$, 
feature vector $\phi_{(i,j)} =w_{i} +w_{j} \in \mathbb{R}^{3406}$ 
where $w_i$ is generated 
as a frequency of words in the statement $i$.
Then, we normalize $\phi_{(i,j)}$. 
The match quality ${ x}_{(i,j)}$ is measured as the cosine similarity between two statements $i$ and $j$,
i.e., ${ x}_{(i,j)} = w_i^\top w_j$, 
and we remove statement pairs $(i,j)$ such that its match quailty 
${ x}_{(i,j)}$ is smaller than $15\%$ of the maximum one.
Finally, by choosing $q_{(i,j)} = \exp\left( 0.01 \cdot {x}_{(i,j)} \right)$,
we obtain $\binom{8}{2}=28$ kernel matrices of dimension $d$ from $516$ to $4,000$. 


Figure \ref{fig:match} reports log-probability ratios and speedups of {Algorithm \ref{alg:batch}} under
the 28 kernels.
We observe that {Algorithm \ref{alg:batch}} looses $0.03$-approximation ratio on average,
compared to  \textsc{Lazy}, under the real-world kernels.
Interestingly, \textsc{Softmax} runs much slower than even \textsc{Lazy},
{
while our algorithm runs faster than \textsc{Lazy} for large dimension, e.g.,
$8$ times faster for $d=4,000$ corresponding to transcripts of Bush and Rubio.
}


\begin{figure}[h]
\begin{center}
\subfigure[]{\includegraphics[width=0.4\textwidth]{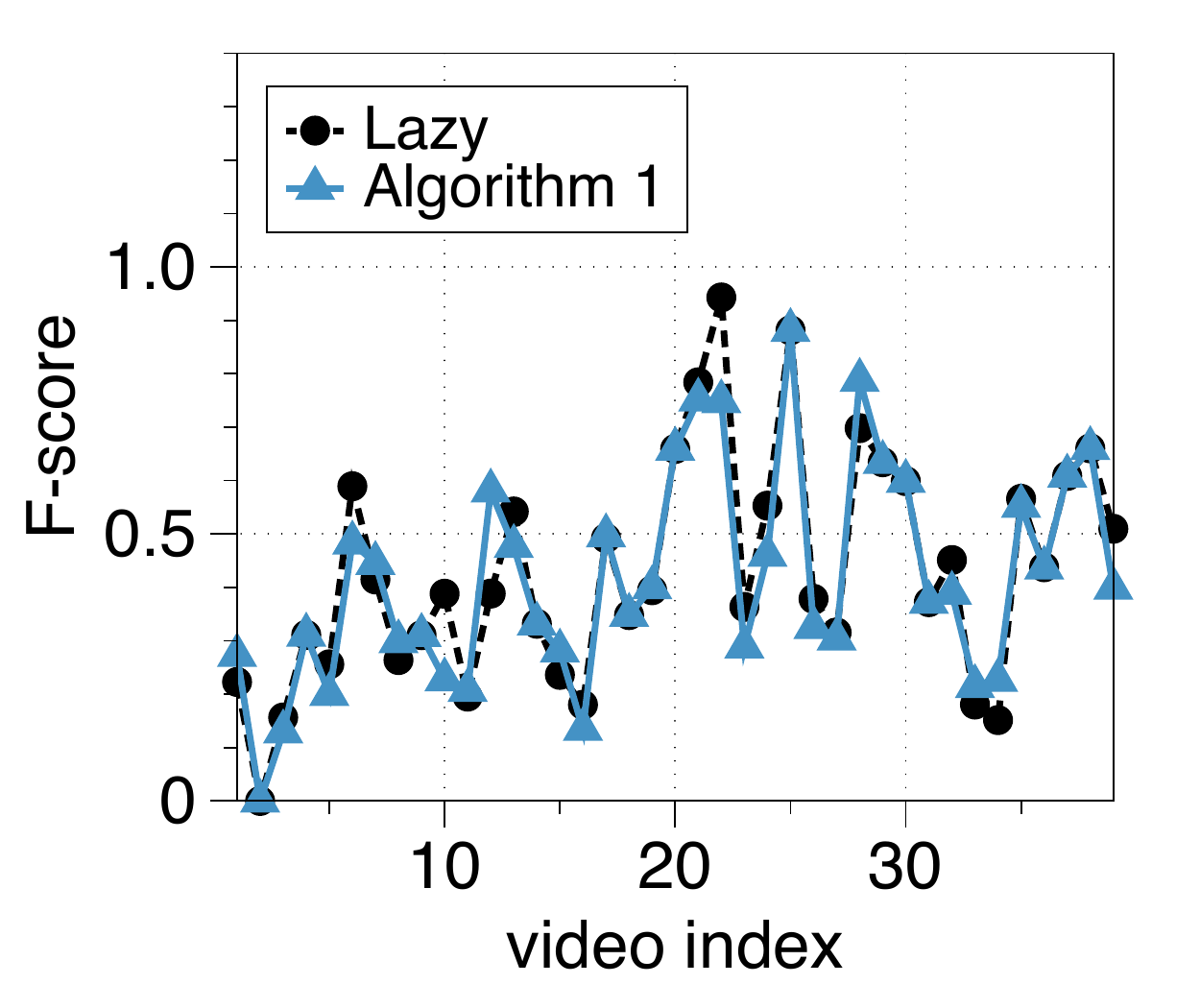} \label{fig:fscore}}
\hspace{-0.15in}
\subfigure[]{\includegraphics[width=0.4\textwidth]{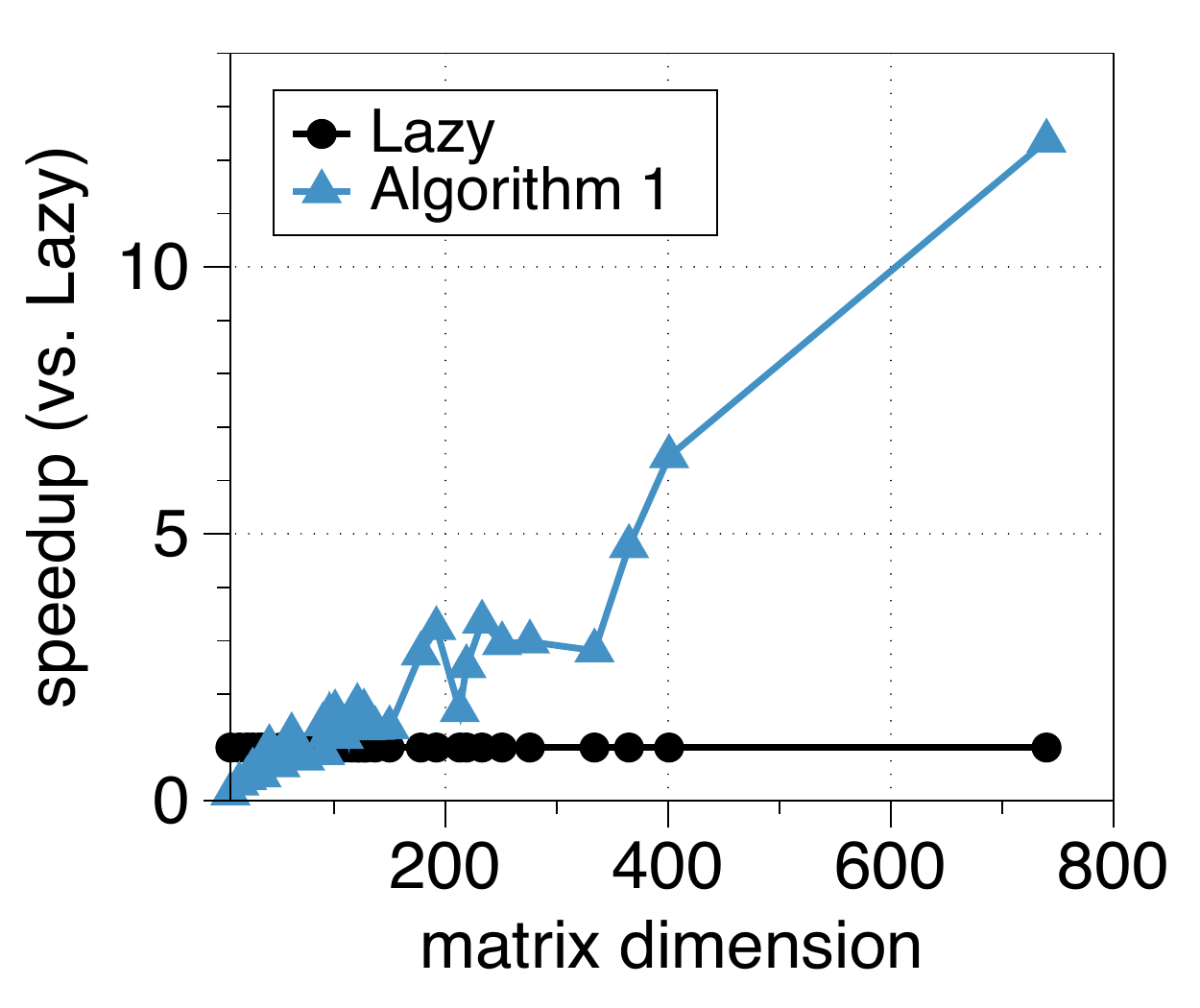} \label{fig:speedupvideo}}
\subfigure[]{\includegraphics[width=0.6\textwidth]{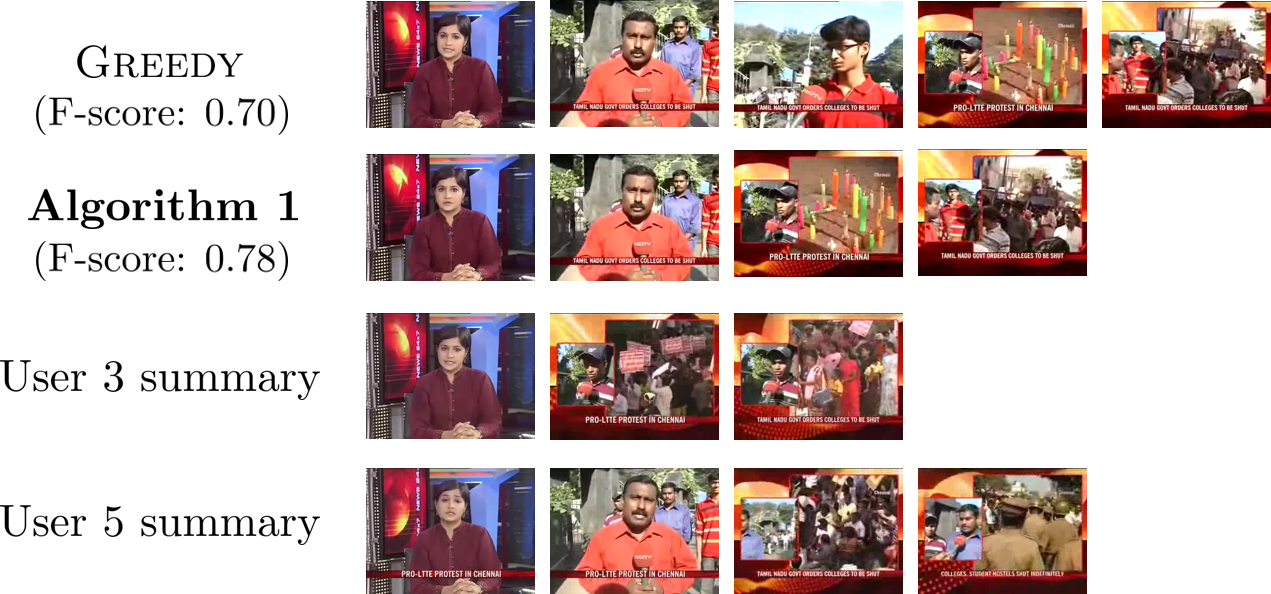} \label{fig:video99}}
\vskip -0.1in
\caption{Plot of (a) F-scores for {Algorithm \ref{alg:glin}} compared to \textsc{Lazy}
and (b) speedup of both algorithms.
(c) shows the summaries of YouTube video of index 99. Images in the first row are summaries produced by \textsc{Lazy} and the second row images illustrate those produced by {Algorithm \ref{alg:glin}}. 
The bottom 2 rows reflect `real' user summaries.} \label{fig:video}
\end{center}
\end{figure}
\noindent {\bf Video summarization.}
We evaluate our proposed algorithms 
video summarization.
We use 39 videos from a Youtube dataset \cite{de2011vsumm},
and the trained DPP kernels from \cite{gong2014diverse}.
Under the kernels, we found that
the numbers of selected elements from algorithms are typically small (less than 10),
and hence we use {Algorithm \ref{alg:glin}} instead of its batch version 
{Algorithm \ref{alg:batch}}.
For performance evaluation, we use an F-score based on five sets of user summaries 
where it measures the quality across two summaries. 

Figure \ref{fig:fscore} illustrates F-score for \textsc{Lazy} and {Algorithm \ref{alg:glin}}
and Figure \ref{fig:speedupvideo} reports its speedup. 
Our algorithm achieves over 13 times speedup in this case,
while it produces F-scores that are very similar to those of \textsc{Lazy}. 
For some video, it achieves even better F-score, as illustrated in \ref{fig:video99}.

\section{Conclusion} \label{sec:conclusion}
We {have presented} fast algorithms for the MAP inference task of large-scale DPPs.
Our main idea is to amortize common determinant computations via
linear algebraic techniques and recent log-determinant approximation methods.
Although we primarily focus on a special matrix optimization,
we expect that several ideas developed in this paper would be useful for other related matrix computational problems,
in particular, involving multiple determinant computations.

\bibliography{biblist}
\bibliographystyle{apalike}

\end{document}